\documentclass{emulateapj}

\shorttitle{AD Leonis: Radial velocity signal}
\shortauthors{Tuomi et al.}

\begin{document}

\title{AD Leonis: Radial velocity signal of stellar rotation or spin-orbit resonance?}

\author{Mikko Tuomi\altaffilmark{1},
Hugh R. A. Jones\altaffilmark{1},
John R. Barnes\altaffilmark{2},
Guillem Anglada-Escud\'e\altaffilmark{3},
R. Paul Butler\altaffilmark{4},
Marcin Kiraga\altaffilmark{5},
Steven S. Vogt\altaffilmark{6}}

\altaffiltext{1}{Centre for Astrophysics Research, School of Physics, Astronomy and Mathematics, University of Hertfordshire, College Lane,\\ AL10 9AB, Hatfield, UK}
\altaffiltext{2}{Department of Physical Sciences, The Open University, Walton Hall, Milton Keynes, MK7 6AA, UK}
\altaffiltext{3}{School of Physics and Astronomy, Queen Mary University of London, 327 Mile End Road, E1 4NS, London, United Kingdom}
\altaffiltext{4}{Department of Terrestrial Magnetism, Carnegie Institution of Washington, 5241 Broad Branch Road NW, Washington D.C. USA\\ 20015-1305}
\altaffiltext{5}{Warsaw University Observatory, Aleje Ujazdowskie 4, 00-478 Warszawa, Poland}
\altaffiltext{6}{UCO/Lick Observatory, University of California, Santa Cruz, CA 95064, USA}

\begin{abstract}
  AD Leonis is a nearby magnetically active M dwarf. We find Doppler variability with a period of 2.23 days as well as photometric signals: (1) a short period signal which is similar to the radial velocity signal albeit with considerable variability; and (2) a long term activity cycle of 4070$\pm$120 days. We examine the short-term photometric signal in the available ASAS and MOST photometry and find that the signal is not consistently present and varies considerably as a function of time. This signal undergoes a phase change of roughly 0.8 rad when considering the first and second halves of the MOST data set which are separated in median time by 3.38 days. In contrast, the Doppler signal is stable in the combined HARPS and HIRES radial velocities for over 4700 days and does not appear to vary in time in amplitude, phase, period or as a function of extracted wavelength. We consider a variety of star-spot scenarios and find it challenging to simultaneously explain the rapidly varying photometric signal and the stable radial velocity signal as being caused by starspots co-rotating on the stellar surface. This suggests that the origin of the Doppler periodicity might be the gravitational tug of a planet orbiting the star in spin-orbit resonance. For such a scenario and no spin-orbit misalignment, the measured $v \sin i$ indicates an inclination angle of 15.5$\pm$2.5 deg and a planetary companion mass of 0.237$\pm$0.047 M$_{\rm Jup}$.
\end{abstract}



\section{Introduction}

The Doppler spectroscopy technique has been a successful method for the detection of planets orbiting nearby stars by enabling observers to measure the changes in stellar radial velocities caused by planets orbiting them on Keplerian orbits. For such detection of planets, M dwarfs are especially fruitful targets by being hosts to at least 2.5 planets per star \citep{dressing2015,tuomi2017}; because their star-planet mass-ratios are lower than for F, G, and K dwarfs and therefore better enable detections of planetary signals; and because they are the most frequent stars in the galaxy and in the Solar neighbourhood \citep{chabrier2000,winters2015}.

Some nearby M dwarfs, such as GJ 581 \citep{vogt2010,vogt2012,tuomi2011,baluev2013,robertson2014}, GJ 667C \citep{anglada2012,anglada2013,feroz2014} and GJ 191 \citep{anglada2014,anglada2016b,robertson2015b} have been sources of controversies in the sense that different authors have interpreted the observed signals differently or even disagreed in how many signals could be detected. However, the controversial signals in the radial velocities of these targets have very low amplitudes, which necessarily makes their detection and interpretation difficult. This is not the case for AD Leonis (AD Leo, GJ 388, BD +20 2465) that has been reported to experience radial velocity variations with a period of 2.23 days \citep{bonfils2013,reiners2013}. Both \citet{bonfils2013} and \citet{reiners2013} interpreted the 2.23-day periodicity as a signal originating from the co-rotation of starspots on the stellar surface because the spectra showed line asymmetries that were correlated with the velocity variations. This means that the radial velocities of AD Leo might provide a benchmark case for examining the differences between Doppler signals caused by stellar rotation and planets on Keplerian orbits.

\citet{newton2016} has articulated the detectability challenge that is faced for older M dwarfs. Their range of stellar rotation periods coincides with both the periods where many of their planets lie as well as with their habitable zones. Therefore, a rotational signal might impersonate a radial velocity signal that would be assigned to a candidate planet. A further consideration is that the rotation period of star might become locked to the orbital period of the planet and that eventually spiral in will occur \citep[e.g.][]{hut1980,adams2015}. For \emph{Kepler} stars there does appear to be a dearth of planets at short orbital periods around fast rotating stars \citep{mcquillan2013,teitler2014}. Only slow stellar rotators, with rotation periods longer than 5-10 days, have planets with periods shorter than 2 or 3 days \citep[see e.g. Fig. 2 of][]{mcquillan2013}. \citet{teitler2014} ran numerical simulations to investigate ``why there is a dearth of close-orbiting planets around fast-orbiting stars'' and find that this can be attributed to tidal ingestion of close-in planets by their host stars. Finding examples of such stars in the Solar neighbourhood would then enable studying this mechanism in detail.

In the current work, we analyse the available HARPS and HIRES data in order to test the validity of the interpretation that the 2.23-day signal in the radial velocities of AD Leo is indeed caused by stellar rotation rather than tidal locking. In particular, we study the properties of the signal given different models accounting for activity-induced radial velocity variations or not, and by examining the dependence of the signal on the spectral wavelength range used to derive the differential radial velocities. We also analyse the \emph{All-Sky Automated Survey} \citep[ASAS;][]{pojmanski1997,pojmanski2002} V-band photometry data and \emph{Microvariability and Oscillations of STars} (MOST) photometry \citep{hunt2012} of the target in order to study the signatures of stellar rotation. Moreover, we attempt to explain the photometric and spectroscopic variability by simulating simple starspot scenarios. Finally, we compare the results to other known rapidly rotating nearby M dwarfs in order to see what, if any, connections there are between photometric rotation periods and radial velocity variations.

\section{AD Leo}

AD Leo (GJ 388) is a frequently flaring \citep{hunt2012,buccino2014} M4.5V dwarf with a parallax of 213$\pm$4 mas implying a distance of only 4.9 pc. Based on \citet{delfosse2000} and the V and J-band magnitudes, \citet{bonfils2013} estimate a mass of 0.42 M$_{\odot}$ and a luminosity of 0.023 L$_{\odot}$ as the star was included in their \emph{HARPS Search for Southern Extra-Solar Planets} programme. Given V and J magnitudes of 9.52 \citep{zacharias2013} and 5.449$\pm$0.027 \citep{cutri2003} we obtain a mass of 0.36 by also applying the relation of \citet{delfosse2000}. According to \citet{houdebine2016} the star has a radius of 0.436$\pm$0.049 R$_{\odot}$ and effective temperature of 3414$\pm$100 K. \citet{neves2012} estimate the metallicity of AD Leo to be [Fe/H] $=$ 0.07 and \citet{royas-ayala2013} give a value of 0.28$\pm$0.17.

AD Leo has been claimed to be a photometrically variable star with a $24 \pm 2$ mmag sinusoidal variability\footnote{To remove ambiguity, we define amplitude such that it denotes parameter $A$ in $f(x) = A\sin(x)$, not $2A$ as in \citet{spiesman1986}.} and photometric periodicity of 2.7$\pm$0.05 days \citep{spiesman1986}. Although the statistical significance of this periodicity was not discussed in \citet{spiesman1986}, they interpret the result as an indication of the co-rotation of starspots on the stellar surface and thus, effectively, the rotation period of the star. Based on spectropolarimetry, \citet{morin2008} reported that the star rotates with a period of 2.2399$\pm$0.0006 days (they also gave alternative solutions at periods of 2.2264 and 2.2537 days).

However, the strongest evidence in favour of the short rotation period of AD Leo comes from MOST satellite's photometric observations. MOST observations were reported to contain strong evidence for a periodicity of 2.23$^{+0.36}_{-0.27}$ days \citep{hunt2012} caused by ``spots distributed at different longitudes or, possibly, that the modulation is caused by varying surface coverage of a large polar spot or a spot that is viewed nearly pole-on.'' This suggests a young age and indeed \citet{shkolnik2009} have estimated an age of 25-300 Myr. Since the results of \citet{hunt2012} were based on a MOST photometric time-series with a baseline of only eight days, and there was evidence for variation of the parameters of the periodic signal caused by stellar rotation, their results suggest the presence of changing spot patterns that would be unlikely to produce stable signals in photometric or spectroscopic data over longer time-scales.

AD Leo has been observed to be variable on longer time-scales as well. \citet{buccino2014} reported an approximately seven-year activity cycle based on ASAS photometry and CASLEO spectroscopy. Although ASAS could not cover a whole period of this cycle, and CASLEO detected it only weakly significantly (with a false alarm probability of 8\%), together they indicate the presence of such a cycle rather convincingly.

It is worth noting that \citet{engle2009} mentioned a photometric periodicity of 2.23 days (with an amplitude of $\approx$ 17 mmag, judging by their phase-folded plot) but the significance and uniqueness of their solution was not discussed.

The estimated $v \sin i$ of AD Leo is 2.63 kms$^{-1}$ \citep{houdebine2016}. Together with a radius estimate of 0.436$\pm$0.049, \citet{houdebine2016} then calculated a projected rotation period of 8.38$^{+1.2}_{-1.1}$ days. Thus, because the rotation period of the star is 2.23 days, this implies inclination of the rotation axis of 15.5$^{\circ}$ $^{+2.5}_{-2.0}$, which means that the star is oriented nearly pole-on.

\section{Spectroscopic and photometric data}

\subsection{HARPS radial velocities}

\begin{figure}
\center
\includegraphics[angle=270, width=0.40\textwidth,clip]{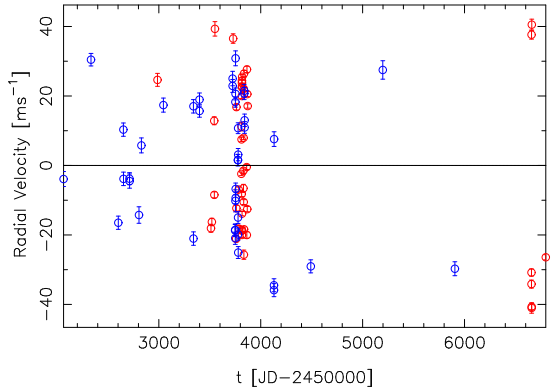}
\caption{HARPS (red) and HIRES (blue) radial velocities of AD Leo with respect to their mean values.}\label{fig:velocities}
\end{figure}

Spectroscopic data of AD Leo were obtained from two sources. We downloaded the publicly available data products of the \emph{High-Accuracy Radial velocity Planet Searcher} \citep[HARPS;][]{mayor2003} from the \emph{European Southern Observatory} archive and processed them with the TERRA algorithms of \citet{anglada2012b}. As a result, we obtained a set of 47 radial velocities with a baseline of 3811 days and root-mean-square (RMS) estimate of 22.59 ms$^{-1}$. Given a mean instrument uncertainty of 0.94 ms$^{-1}$, there are variations in the data that cannot be explained by instrument noise alone. The majority (28) of these were obtained over a short period of 76 days between JDs 2453809 and 2453871 enabling the detection of the signal at a period of 2.23 days \citep{bonfils2013,reiners2013}. The HARPS radial velocities of AD Leo are shown in Fig. \ref{fig:velocities} for visual inspection.

\subsection{HIRES radial velocities}

The second set of spectroscopic data was obtained by the \emph{High Resolution Echelle Spectrometer} \citep[HIRES;][]{vogt1994} of the Keck I telescope. This set of 42 radial velocities has a baseline of 3841 days and an RMS of 18.93 ms$^{-1}$. With an average instrument uncertainty of 1.90 ms$^{-1}$, this data also indicates excess variability that cannot be explained by pure instrument noise. HIRES velocities of AD Leo are also shown in Fig. \ref{fig:velocities} as published in \citet{butler2017}.

\subsection{ASAS photometry}

\begin{figure}
\center
\includegraphics[angle=270, width=0.40\textwidth,clip]{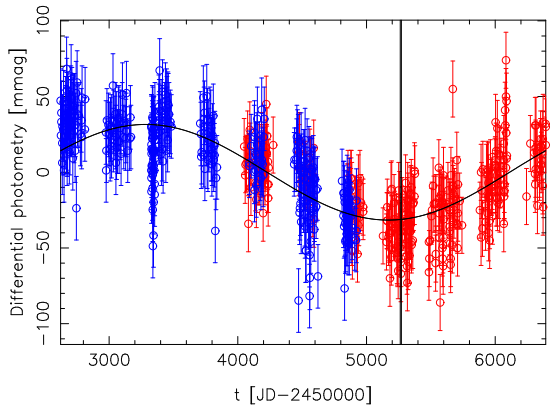}
\caption{ASAS-North (red) and ASAS-South (blue) V-band photometry data of AD Leo with respect to the data mean. Only 'Grade A' is shown with all 5-$\sigma$ outliers removed. An offset of 25.9 mmag has been accounted for. The uncertainties represent estimated excess variability in the data. The black curve denotes the long-period activity-cycle of the star reported by \citep{buccino2014}. The position of MOST observing run is denoted by a vertical line.}\label{fig:asas_data}
\end{figure}

To study the photometric variability of AD Leo, we obtained ASAS \citep{pojmanski1997,pojmanski2002} V-band photometry data\footnote{\texttt{www.astrouw.edu.pl/asas}} from both ASAS-North (ASAS-N) and ASAS-South (ASAS-S) telescopes. We only selected the 'Grade A' data from the set, removed all 5-$\sigma$ outliers, and obtained sets of 316 and 319 photometric measurements with baselines of 2344 and 2299 days, respectively. The apertures with the least amount of variability showed brightnesses of 9263.8$\pm$25.2 and 9333.0$\pm$28.3 mmag. These values are somewhat different and we thus only compared the two time-series by assuming an offset that was a free parameter of the model. The ASAS data is shown in Fig. \ref{fig:asas_data} together with estimated long-period cycle that was clearly present in the data as also observed by \citet{buccino2014}. We note that \citet{kiraga2007} did not discuss AD Leo when publishing photometric rotation periods of nearby M dwarfs.

\subsection{MOST photometry}

\begin{figure}
\center
\includegraphics[angle=270, width=0.40\textwidth,clip]{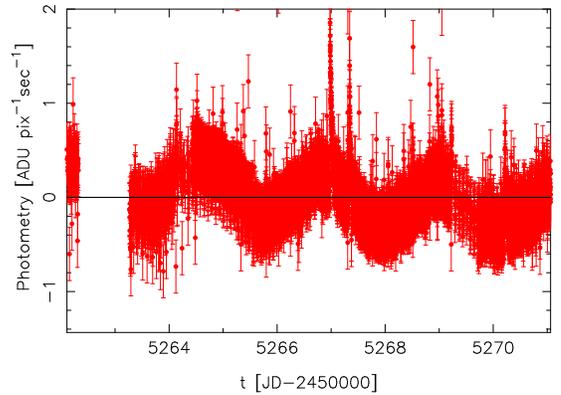}
\caption{Raw MOST photometry data of AD Leo with respect to the data mean \citep[see][Fig. 1]{hunt2012}. The solid horizontal line represents the data mean.}\label{fig:most}
\end{figure}

We also obtained the raw MOST photometry data\footnote{MOST photometry data was kindly provided by Nicholas M. Hunt-Walker.} as discussed in \citet{hunt2012}. The data is presented in Fig. \ref{fig:most} for visual inspection and consists of 8592 individual observations over a baseline of roughly 9 days. The MOST observing run is denoted in Fig. \ref{fig:asas_data} by a vertical line.

\section{Photometric variability of AD Leo}\label{sec:asas}

As observed by \citet{buccino2014}, the dominant feature in the ASAS-S data is a long-period signal caused by the star's activity cycle. In combination with ASAS-N data, we estimate this cycle to have a period of 4070$\pm$120 days, when using the standard deviation of the period parameter to describe the uncertainty. The 99\% credibility interval of this period is [3730, 4450] days.

Modelling the long-period signal, we analysed the ASAS photometry of AD Leo by calculating likelihood-ratio periodograms of both ASAS-N and ASAS-S data sets. These periodograms were calculated by assuming the measurements were independent and normally distributed. However, instead of the common Lomb-Scargle periodogram \citep{lomb1976,scargle1982} that is equivalent to the minimisation of $||C^{1/2}[m - f(\theta)]||^{2}$, where $m$ is the measurement vector, $C$ is the covariance matrix, and the model is defined as $f(\theta) = a_{1} \sin \omega t + a_{2} \cos \omega t$, we also included a second order polynomial such that $f(\theta) = a_{1} \sin \omega t + a_{2} \cos \omega t + a_{3} + a_{4} t + a_{5} t^{2}$ \citep[see also][]{butler2017} that thus accounted for the long-period cycle seen as a second order curvature in both ASAS-N and ASAS-S data sets in Fig. \ref{fig:asas_data}. In our notation, $\theta = (a_{1}, ..., a_{5})$ represents the parameter vector, $\omega$ is the frequency, and $t$ is time.

With the model containing a sinusoidal signal and the polynomial terms, we then attempted detecting signals by looking at the likelihood-ratios of models with and without the sinusoid (or Keplerian function, when analysing radial velocities). Our signal detection criteria were such that i) signal improved the model significantly; ii) the signal was unique and well-constrained in the period space; and iii) the amplitude parameter of the signal was well-constrained such that it was statistically significantly different from zero \citep[see e.g.][]{tuomi2012}. The significance of the signal was determined by calculating the likelihood-ratio $L_{\rm r}$ of models with and without signals, and by seeing if $\ln L_{\rm r} > \alpha$, where threshold $\alpha$ was set equal to 16.27 for sinusoidal signal (three free parameters) and 20.52 for Keplerian signals (five free parameters), i.e. such that false alarm probability (FAP) was less than 0.1\%. We also calculated whether the signals exceeded detection thresholds such that the model probabilities roughly estimated with the Bayesian information criterion (BIC) increased by a factor of 150 \citep{kass1995,feng2016}.

\begin{figure}
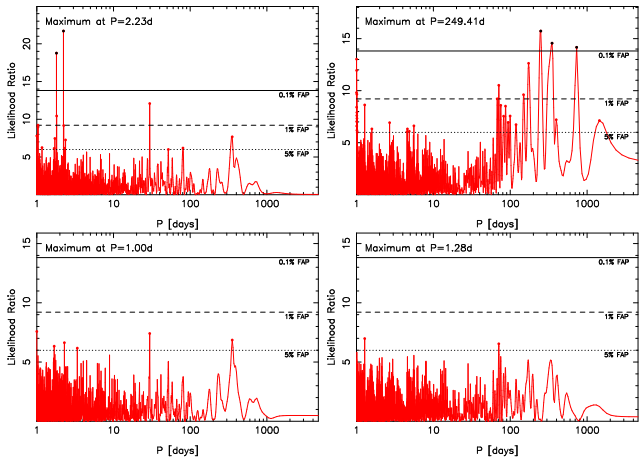

\center
\includegraphics[angle=270, width=0.23\textwidth,clip]{GJ388_01_mlp_ASAS-N_logp.ps}
\includegraphics[angle=270, width=0.23\textwidth,clip]{GJ388_00_mlp_ASAS-S_logp.ps}

\includegraphics[angle=270, width=0.23\textwidth,clip]{GJ388_02_mlp_ASAS-N_logp.ps}
\includegraphics[angle=270, width=0.23\textwidth,clip]{GJ388_01_mlp_ASAS-S_logp.ps}
\caption{Logarithm of the likelihood-ratio periodogram of the ASAS-N (left panels) and ASAS-S (right panels) V-band photometry of AD Leo when accounting for the long-period variability seen in Fig. \ref{fig:asas_data}. The bottom panel shows the residual periodogram after subtracting the strongest periodic signal. The red (black) filled circles denote that maxima exceeding the 5\% (0.1\%) FAP threshold.}\label{fig:asas_periodogram}
\end{figure}

The likelihood-ratio periodogram\footnote{We note that the periodograms are calculated for a model with only two additional free parameters because the period parameter is fixed and the 0.1\%, 1\% and 5\% FAP thresholds values are thus at 13.82, 9.21, and 5.99 in the periodograms.} of ASAS-N data shows a maximum in excess of the 0.1\% FAP at a period of 2.22791 [2.22736, 2.22857] days with an amplitude of 9.3 [5.2, 13.0], where the uncertainties have been presented as 99\% credibility intervals. This periodicity corresponds to the rotation period of AD Leo. No other periodicities could be found in the ASAS-N photometry. Instead, the periodogram of ASAS-S data shows evidence for another period of 257.9 [250.2, 262.5] days with an amplitude of 6.7 [3.0, 11.3] mmag (Fig. \ref{fig:asas_periodogram}). The secondary periodogram maxima exceeding the 0.1\% FAP threshold at periods of 350 and 730 days are likely caused by annual gaps in the data (see Fig. \ref{fig:asas_data}) and aliasing. After accounting for the long-period cycle, there were no periodicities in the combined ASAS data (allowing an offset between the data sets) exceeding the 0.1\% FAP, although four periods -- 1.81, 2.23, 376.1, and 426.8 days -- had likelihood ratios exceeding the 1\% FAP threshold.

As can be seen in Fig. \ref{fig:asas_periodogram}, there is only evidence for the photometric rotation period in the ASAS-N data. This signal is not present in ASAS-S data at all. In particular, the photometric signal at a period of 2.23 days with an amplitude of 9.3 mmag should have been clearly visible in the ASAS-S data considering that a much longer periodicity of 257.9 days with an amplitude of only 6.7 mmag could be confidently detected. This suggests that when the star reached the brightness minimum corresponding to the long-period activity cycle (Fig. \ref{fig:asas_data}), the rotation period could not be seen in low-cadence observations such as those obtained by ASAS-S. The MOST observations were taken between JDs 2455262.0 and 2455271.0, which corresponds to the brightness maximum as determined by the ASAS data.

We also analysed separately each of the seven observing seasons (see Fig. \ref{fig:asas_data}) in order to find periodic signals in them. According to the likelihood-ratio periodograms, there was no evidence for strong periodic signals in excess of 1\% FAP in any of the observing seasons. It thus appears evident that although the photometric rotation signal is not strong in data of any given season, the lack of it in the ASAS-S data suggests it is not very stable, likely due to the fact that spot-patterns on the stellar surface differ markedly between different phases of the activity cycle of the star.

\subsection{MOST high-cadence photometry}\label{sec:most}

As reported by \citet{hunt2012}, the MOST high-cadence photometry showed clear evidence in favour of a photometric rotation period of 2.23 days, as well as several flare-events (Fig. \ref{fig:most}). In an attempt to obtain constraints for the variability of the photometric signal in the MOST data, we split the raw data ($N = 8592$) in half -- we then analysed the two sets to quantify any changes in the phase, period, and amplitude of the photometric signal over the MOST baseline of 8.95 days.

According to our results, the phase of a sinusoid changed from 1.06$\pm$0.08 to 1.85$\pm$0.10 rad (when using standard $1\sigma$ uncertainty estimates) from the first half of the data to the second. Variability was also detected in the photometric period and amplitude that changed from 2.289$\pm$0.019 to 2.145$\pm$0.011 days and from 0.2609$\pm$0.0046 to 0.2242$\pm$0.0037 ADU pix$^{-1}$sec$^{-1}$, respectively. Given that the median times of the two MOST data halves differ by only 3.38 days, this evolution takes place on a time-scale comparable to the star's rotation period. It is thus evident that the spot-patterns and active/inactive areas on the star's surface giving rise to the clear photometric rotation signal experience rapid evolution and change considerably over a period of only few days.

The variability of the photometric signal on short time-scales indicates that the signal detected in ASAS-N data is only an average over a baseline of roughly 3000 days. This implies that the photometric amplitude of the signal detected in ASAS-N appears lower than it actually is because it corresponds to an average over different phases in the star's activity cycle. This interpretation is supported by \citet{spiesman1986}, who estimated the amplitude to be $12$ mmag.

\section{Spectroscopic variability of AD Leo}

The radial velocities of AD Leo were found to vary periodically as also observed for the HARPS data by \citet{bonfils2013} and \citet{reiners2013}. We analysed the data by applying the delayed-rejection adaptive-Metropolis (DRAM) algorithm \citep{haario2001,haario2006} that is a generalisation of the Metropolis-Hastings Markov chain Monte Carlo posterior sampling technique \citep{metropolis1953,hastings1970}. This technique has been used to find periodicities in radial velocity data in e.g. \citet{jenkins2014}, \citet{tuomi2014} and \citet{butler2017}.

\subsection{Activity indices}

We also obtained and analysed selected HARPS and HIRES activity indicators -- the S-index measuring the emission of CaII H\&K lines for HARPS and HIRES with respect to the continuum, and the line bisector span (BIS) and full-width at half-maximum (FWHM) for HARPS. 

\begin{figure}
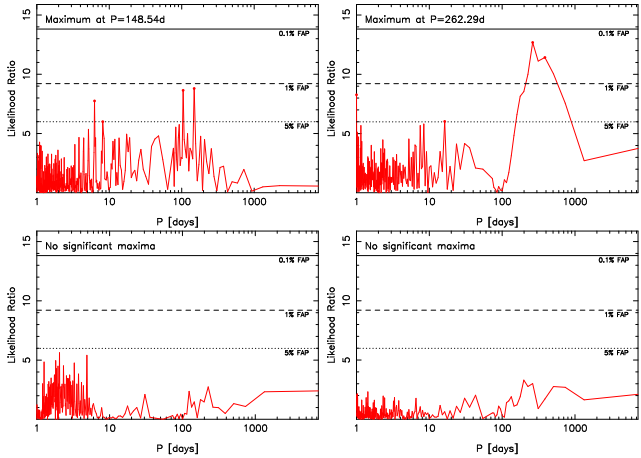

\center
\includegraphics[angle=270, width=0.23\textwidth]{GL388_mlp_HIRES_S_logp.ps}
\includegraphics[angle=270, width=0.23\textwidth]{GJ388_mlp_HARPS_S_logp.ps}

\includegraphics[angle=270, width=0.23\textwidth]{GJ388_mlp_HARPS_BIS_logp.ps}
\includegraphics[angle=270, width=0.23\textwidth]{GJ388_mlp_HARPS_FWHM_logp.ps}
\caption{Logarithm of the likelihood-ratio periodogram of the HIRES S-index (top left panel), HARPS S-index (top right panel), BIS (bottom left panel), and FWHM (bottom right panel).}\label{fig:activity}
\end{figure}

The likelihood-ratio periodogram of the HIRES S-index showed some evidence (in excess of 5\% but not 1\% FAP) for a 150-day periodicity (Fig. \ref{fig:activity}, top panel). This variability could be connected to photometric variability detected in the ASAS-S photometry. We also detect a broad maximum in the periodogram of HARPS S-indices at a period of 300 days (Fig. \ref{fig:activity}, second panel). However, we could not observe any evidence for periodicities in the HARPS BIS and FWHM values (Fig. \ref{fig:activity}). This result is consistent with that of \citet{reiners2013} who discussed hints of evidence for periodicities in the HARPS BIS values but the corresponding periodic signals only barely exceeded 5\% FAP in their analyses.

\subsection{Radial velocities}

We modelled the radial velocities by accounting for a Keplerian signal ($f_{k}$), reference velocity $\gamma_{l}$ of instrument $l$, a linear trend ($\dot{\gamma}$), linear dependence of the velocities on the activity indices $\xi_{i,j,l}$ with parameter $c_{j,l}$, and moving average component with exponential smoothing accounting for some of the red features \citep[e.g.][]{baluev2009,tuomi2014} in the radial velocity noise. The statistical model for a measurement $m_{i,l}$ at epoch $t_{i,l}$ is thus
\begin{eqnarray}\label{eq:model}
  m_{i,l} = f_{k}(t_{i,l}) + \gamma_{l} + \dot{\gamma} + \sum_{j} c_{j,l} \xi_{i,j,l} \nonumber\\
  + \phi_{l} \exp \bigg\{ \frac{t_{i-1}-t_{i}}{\tau} \bigg\}r_{i-1,l} + \epsilon_{i,l} ,
\end{eqnarray}
where $\tau$ was set equal to 4 days because we expect to see correlations on that time-scale but not on longer time-scales \citep{tuomi2014} and $r_{i,l}$ represents the residual after subtracting the deterministic part of the model from the data. The Gaussian random variable $\epsilon_{i,l}$ represents the white noise in the data -- it has a zero mean and a variance of $\sigma_{i}^{2} + \sigma_{l}^{2}$ where $\sigma_{l}$ is a free parameter for all instruments quantifying the excess white noise, or ``jitter'', in the data.

\begin{figure}
\center
\includegraphics[angle=270, width=0.40\textwidth]{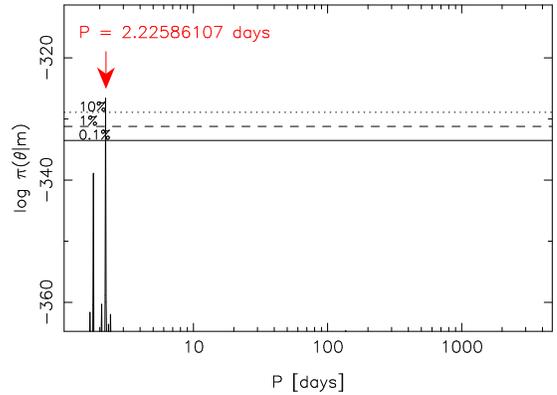}
\caption{Estimated posterior density as a function of the signal period given the HARPS and HIRES radial velocities of AD Leo. The horizontal lines represent the 10\% (dotted), 1\% (dashed), and 0.1\% (solid) equiprobability thresholds with respect to the global maximum denoted by the red arrow.}\label{fig:posterior_period}
\end{figure}

\begin{figure}
\center
\includegraphics[angle=270, width=0.40\textwidth]{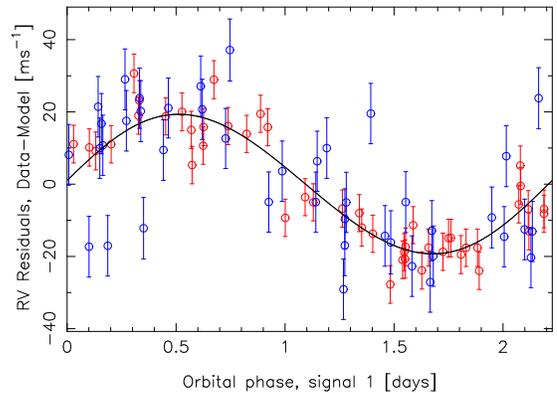}
\caption{HARPS (red) and HIRES (blue) radial velocities of AD Leo folded on the phase of the signal.}\label{fig:signal}
\end{figure}

The combined HARPS and HIRES radial velocities of AD Leo contained a periodic signal that was clearly identified by our DRAM samplings of the parameter space (Fig. \ref{fig:posterior_period}). We show the HARPS and HIRES radial velocities folded on the phase of the signal in Fig. \ref{fig:signal} for visual inspection. This signal was supported by both data sets -- the maximised log-likelihood (natural logarithm) increased from -199.6 to -179.1 for HIRES and from -196.6 to -153.7 for HARPS exceeding any reasonable statistical significance thresholds\footnote{For a Keplerian signal with five free parameters, a natural logarithm of a likelihood-ratio of 20.52 corresponds to a 0.1\% false alarm probability.}. Using the BIC to determine the significance of the signal \citep[see][]{schwarz1978,feng2016}, we obtain an estimate for the logarithm of Bayes factor in favour of a model with one signal of 52.25 -- considerably in excess of the detection threshold of 5.01 corresponding to a situation where the model is 150 times more probable \citep[e.g.][]{kass1995}. This signal satisfied all the signal detection criteria of \citet{tuomi2012}, i.e. in addition to satisfying the significance criterion, it corresponded to a unique posterior probability maximum constrained from above and below in the period and amplitude spaces.

As also reported by \citet{bonfils2013} and \citet{reiners2013}, we observed a negative correlation between HARPS radial velocities and BIS values. The linear parameter $c_{j,l}$ (see the model above) that quantifies this dependence of the HARPS velocities on the observed BIS values was found to have a value of -2.12 [-2.96, -1.30], which is significantly different from zero at a 7.7$\sigma$ level. Similar strong correlations were not found with HARPS FWHM and S-index. There was a weaker correlation between the HIRES velocities and S-index with a parameter value of 1.62 [0.21, 3.18] ms$^{-1}$, which indicates a 3.1$\sigma$ significance.

We have tabulated the parameters of the radial velocity signal, when assuming it has a Keplerian shape, in Table \ref{tab:GJ388_solution} together with the ``nuisance parameters'' in the statistical model. We also tabulated the estimates when not including the correlations between velocities and the activity indicators in the model -- i.e. when fixing parameters $c_{i,l} = 0$ for all indices and both instruments. We note that the two solutions are not statistically significantly different for the Keplerian parameters -- this is demonstrated by the fact that the 99\% credibility intervals of the parameters of the signal are not distinct between the two models\footnote{This means that when considering the intervals to be Bayesian credibility sets, these sets have an intersection that is not an empty set.}. This implies that the properties of the signal are independent of whether we account for the correlations between the velocities and activity data or not. However, as discussed above, there is a significant correlation between the HARPS radial velocities and BIS values and we thus consider the solution that includes the correlations to be more trustworthy (the solution on the left hand side in Table \ref{tab:GJ388_solution}).

\begin{table*}
\begin{minipage}{\textwidth}
\begin{center}
\caption{Maximum \emph{a posteriori} estimates and 99\% credibility intervals of the parameters of a model with one Keplerian signal and with or without correlations with the activity indicators. Parameters $K$, $P$, $e$, $\omega$, and $M_{0}$ are the Keplerian parameters: radial velocity amplitude, signal period, eccentricity, argument of periapsis, and mean anomaly with respect to epoch $t = 2450000$ JD, respectively. The uncertainty in stellar mass has been accounted for when estimating the minimum mass and semi-major axis corresponding to the signal under planetary interpretation for its origin.}\label{tab:GJ388_solution}
\begin{tabular}{lccccc}
\hline \hline
Parameter & Full model & No activity correlations \\
\hline
$K$ [ms$^{-1}$] & 19.11 [15.69, 22.54] & 23.18 [19.68, 26.67] & \\
$P$ [days] & 2.22579 [2.22556, 2.22593] & 2.22579 [2.22566, 2.22592] & \\
$e$ & 0.015 [0, 0.147] & 0.028 [0, 0.161] & \\
$\omega$ [rad] & 4.7 [0, 2$\pi$] & 5.1 [0, 2$\pi$] & \\
$M_{0}$ [rad] & 6.2 [0, 2$\pi$] & 0.3 [0, 2$\pi$] & \\
$a$ [AU] & 0.024 [0.021, 0.026] & 0.024 [0.021, 0.027] & \\
$m \sin i$ [M$_{\oplus}$] & 19.7 [14.6, 25.3] & 23.8 [17.9, 29.8] & \\
\hline
$\dot{\gamma}$ [ms$^{-1}$year$^{-1}$] & -1.50 [-3.00, -0.31] & -1.68 [-2.46, -0.89] & \\
$\sigma_{\rm HARPS}$ [ms$^{-1}$] & 5.52 [4.27, 6.90] & 6.58 [5.36, 8.09] & \\
$\sigma_{\rm HIRES}$ [ms$^{-1}$] & 8.40 [7.07, 9.88] & 8.74 [7.47, 10.31] & \\
$\phi_{\rm HARPS}$ & -0.10 [-0.94, 0.80] & -0.66 [-1, 0.07] & \\
$\phi_{\rm HIRES}$ & 0.61 [0.05, 1] & 0.27 [-0.39, 0.88] & \\
$c_{\rm BIS,HARPS}$ & -2.12 [-2.96, -1.29] & -- & \\
$c_{\rm FWHM,HARPS}$ & 0.230 [-0.141, 0.600] & -- & \\
$c_{\rm S,HARPS}$ [ms$^{-1}$] & -0.14 [-2.82, 2.55] & -- & \\
$c_{\rm S,HIRES}$ [ms$^{-1}$] & 1.62 [0.21, 3.18] & -- & \\
\hline \hline
\end{tabular}
\end{center}
\end{minipage}
\end{table*}

\subsection{Colour- and time-invariance of the radial velocity variations}\label{sec:colour-time}

\begin{figure}
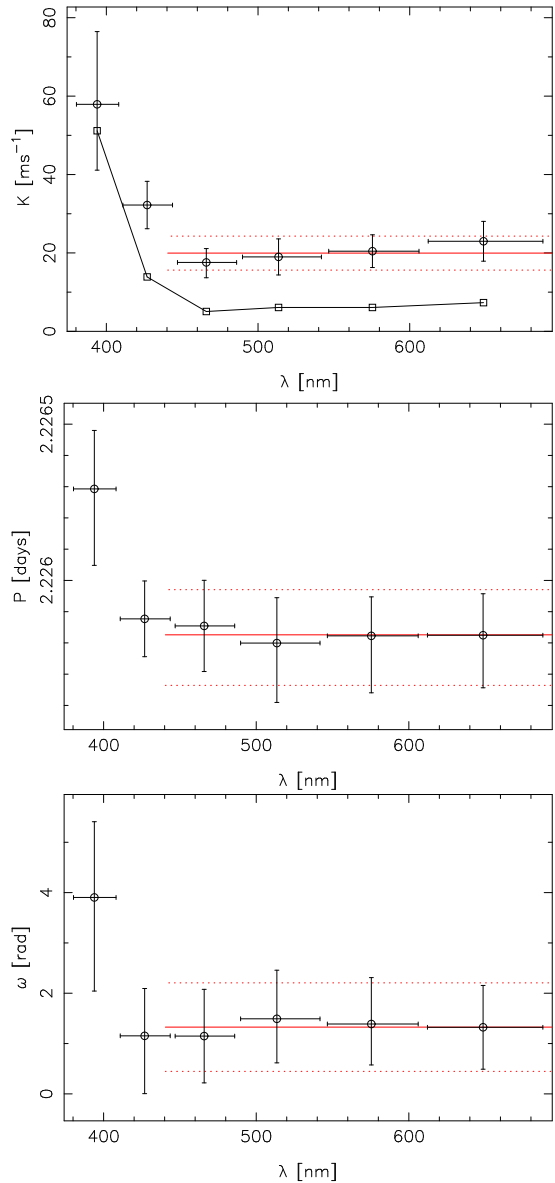

\center
\includegraphics[angle=270, width=0.40\textwidth]{GJ388_amplitude.ps}

\includegraphics[angle=270, width=0.40\textwidth]{GJ388_period.ps}

\includegraphics[angle=270, width=0.40\textwidth]{GJ388_phase.ps}
\caption{Wavelength dependence of the signal in the HARPS radial velocities. Parameter estimates for the velocities calculated for six sets of twelve HARPS orders -- radial velocity amplitude (top panel), signal period (middle panel), and signal phase when assuming a circular solution (bottom panel). The solid (dotted) red horizontal line denotes the estimate (99\% credibility interval) obtained for the HARPS-TERRA velocities based on the reddest 50 orders \citep[see][]{anglada2012b}. In top panel, the squares (and the black lines) denote the data RMS for comparison. The horizontal error bars denote the wavelength range of the twelve HARPS orders used to obtain the radial velocities.}\label{fig:wavelength_dependence}
\end{figure}

A true Keplerian Doppler signal of planetary origin cannot depend on the wavelength range of the spectrograph. Although \citet{reiners2013} tested the wavelength dependence of the signal in the AD Leo velocities by measuring the amplitude of the signal for different subsets of the 72 HARPS orders, they did not account for red noise or correlations between the velocities and activity indicators. We repeated this experiment by calculating the weighted mean velocities for six sets of twelve orders. Apart from the bluest 24 HARPS orders that were found to be heavily contaminated by activity \citep[see also][]{anglada2012b}, we found the parameters of the signal to be independent of the selected wavelength range. This is demonstrated in Fig. \ref{fig:wavelength_dependence} where we have plotted the signal amplitude, period, and phase (when assuming circular solution) as a function of wavelength. The remarkable stability of the signal represents the hallmarks of a Keplerian signal of planetary origin and is difficult to interpret as a signal that is caused by starspots co-rotating on the stellar surface. Although the bluest orders do not appear to agree with the rest of the orders in this respect, we note that the solution for the first twelve orders is actually a highly eccentric solution that arises from the activity induced variations (and correspondingly higher root-mean-square) at the bluest wavelengths.

The standard deviation of the radial velocities from the bluest twelve orders was found to be 77.89 ms$^{-1}$ whereas that of the next bluest twelve orders was 31.07 ms$^{-1}$. This arises due to a combination of lower signal-to-noise as well as the bluer orders being more prone to activity-induced variability. As a consequence, it was not possible for us to detect the signal correctly in the bluest twelve orders, and probably caused biases in the estimated parameters of the signal for the second bluest twelve orders (Fig. \ref{fig:wavelength_dependence}). However, when using the so-called differential velocities of \citet{feng2017} as activity proxies, we could see the signal as a clear probability maximum at the same period in the period space for all six wavelength ranges. Although independent detection of the signal was only possible for the four redmost sets of twelve HARPS orders, the differential velocities helped removing activity-induced variability such that the signal could be seen throughout the HARPS wavelength range.

\begin{figure}
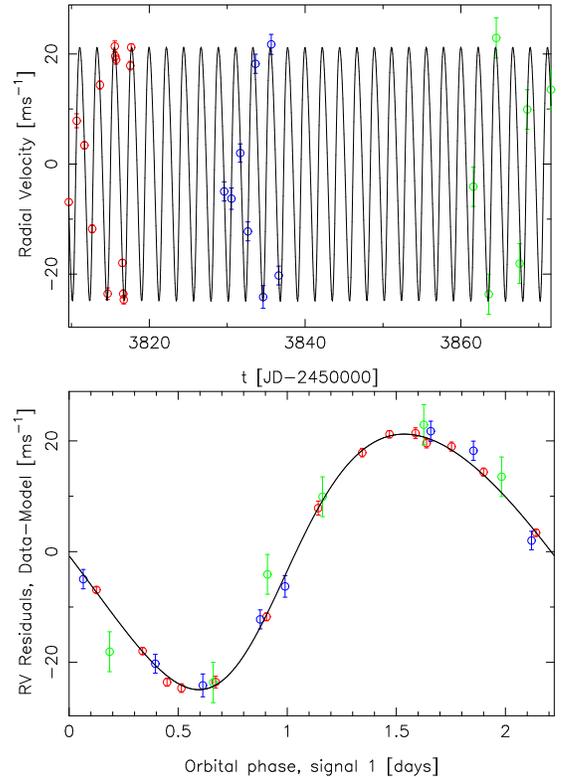

\center
\includegraphics[angle=270, width=0.40\textwidth]{rv_GJ388_01_curvec_COMBINED.ps}

\includegraphics[angle=270, width=0.40\textwidth]{rv_GJ388_01_scresidc_COMBINED_1_HHC.ps}
\caption{HARPS radial velocities of AD Leo between JDs 2453809.7 and 2453871.6 as a function of time (top panel) and folded on the phase of the signal (bottom panel). The black curve denotes the estimated Keplerian curve. The three temporal subsets of data denoted by different colours have been treated as independent data sets with independent nuisance parameters.}\label{fig:zoom_data}
\end{figure}

Another sign-post that a periodic signal in radial velocities is a Keplerian one, caused by a planet orbiting the star, is time-invariance. We tested this time-invariance by looking more closely at a 60-day period during which HARPS was used to observe AD Leo 28 times: over nine ($N = 14$) and eight ($N = 8$) consecutive nights and then six times over a period of eleven nights. These observations are treated as independent data subsets and plotted in Fig. \ref{fig:zoom_data} together with the estimated Keplerian curve. As can be seen in Fig. \ref{fig:zoom_data}, over this period of 60 days, the periodic variability is remarkably stable and consistent with a time-invariant signal.

Due to the lack of another period of observations with suitably high observational cadence, we then analysed in combination the HARPS data not included in the above 60-day period and HIRES data. This combined data set with 19 HARPS and 42 HIRES radial velocities showed evidence for a consistent periodicity with the 60-day HARPS observing run. Assuming zero eccentricity, the signals in the 60-day observing run and the rest of the data have amplitudes of 23.13 [19.28, 26.24] and 18.78 [11.68, 25.87] ms$^{-1}$, periods of 2.22290 [2.21810, 2.22645] and 2.22576 [2.22507, 2.22601] days, and phases of 4.08 [0, 2$\pi$] and 1.83 [0.28, 4.81] rad, respectively. Although the solution given the 60-day HARPS observing run is rather uncertain (the former solution above), this demonstrates that the signal is stable with a precision of almost two orders of magnitude better than was observed for the photometric rotation signal in MOST data in Section \ref{sec:most}.

\subsection{Simulated variable signals}

To investigate whether starspots co-rotating on the stellar surface could produce coherent radial velocity signals over a period of several years, such as the baseline of the radial velocity data of AD Leo of 4733 days, we generated artificial radial velocities to study the detectability of evolving signals in the available radial velocity data.

We generated artificial data sets with the same properties as were observed for the HARPS and HIRES data (Table \ref{tab:GJ388_solution}) and added one sinusoidal signal in the data with $P = 2.23$ days and $K = 20.0$ ms$^{-1}$. Because the MOST photometry indicated that, within a period of nine days, the photometric signal evolved considerably in phase, period and amplitude, we changed the phase of the signal by an angle of $\psi t$, where $\psi$ is a constant and $t$ is time. Parameter $\psi$ was selected to have values $\psi \in \frac{1}{10} [0, 1, ..., 7]$ rad/week -- i.e., the signal was made to vary linearly as a function of time from 0 to 0.7 rad per week. We denote these datasets as S1, S2, ..., S8, respectively. Although only a toy model, this still enabled us to study the sensitivity of our signal detection to evolving signals. To avoid analysing the artificial data with the same model as was used to generate it \citep[i.e. committing an ``inverse crime'';][]{kaipio2005}, we used a noise model with third-order moving average terms with second and third order components fixed such that $\phi_{2} = 0.3$ and $\phi_{3} = 0.1$, respectively, whereas we only applied the first-order moving average model when analysing the data.

\begin{figure*}
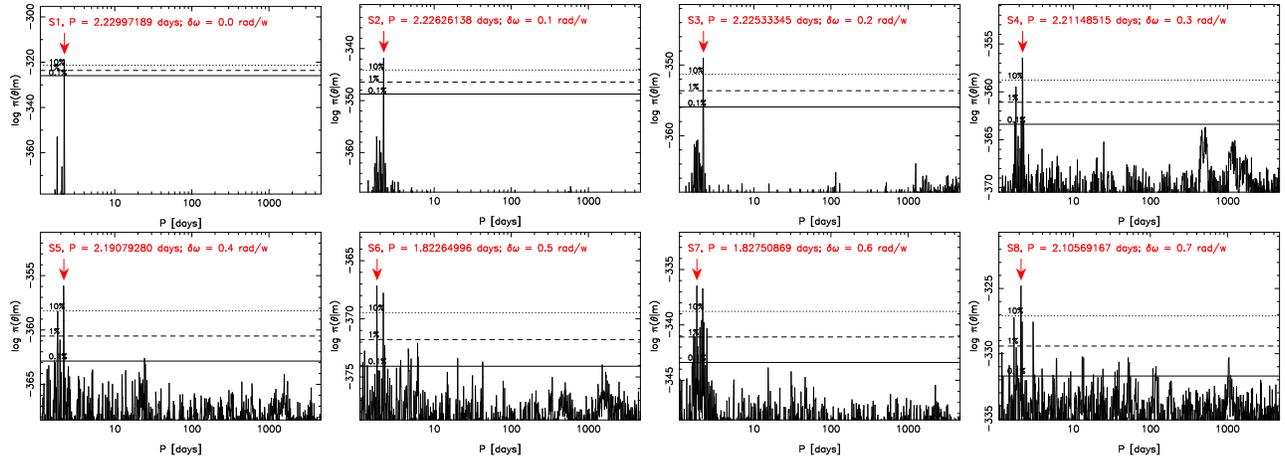

\center
\includegraphics[angle=270, width=0.23\textwidth]{rv_GJ388_R1_01_pcurve_b.ps}
\includegraphics[angle=270, width=0.23\textwidth]{rv_GJ388_R2_01_pcurve_b.ps}
\includegraphics[angle=270, width=0.23\textwidth]{rv_GJ388_R3_01_pcurve_b.ps}
\includegraphics[angle=270, width=0.23\textwidth]{rv_GJ388_R4_01_pcurve_b.ps}

\includegraphics[angle=270, width=0.23\textwidth]{rv_GJ388_R5_01_pcurve_b.ps}
\includegraphics[angle=270, width=0.23\textwidth]{rv_GJ388_R6_01_pcurve_b.ps}
\includegraphics[angle=270, width=0.23\textwidth]{rv_GJ388_R7_01_pcurve_b.ps}
\includegraphics[angle=270, width=0.23\textwidth]{rv_GJ388_R8_01_pcurve_b.ps}
\caption{Estimated posterior probability densities as functions of signal period for artificial data sets with injected signals. The rate of change in the phase of the signals ($\delta \omega$) is denoted in each panel in radians per week. The red arrows denote the global probability maxima and the horizontal thresholds indicate the 10\% (dotted), 1\% (dashed), and 0.1\% (solid) equiprobability thresholds with respect to the maxima.}\label{fig:evolved_phase}
\end{figure*}

The simulated radial velocity data sets were generated such that the injected signal was varied less rapidly than the photometric one in the MOST data and did not disappear contrary to what appears to be the case for the photometric signal in the ASAS data (Fig. \ref{fig:asas_periodogram}). Although we expected a slightly variable signal to cause a clear probability maximum due to the concentration of HARPS data on a period of 60 days (Section  \ref{sec:colour-time}), it was also expected that a more rapidly varying signal would make the detection less probable or impossible. This is indeed what happened, as can be seen in Fig. \ref{fig:evolved_phase} where we have plotted the posterior probability densities as functions of signal periods given artificial datasets with signals whose phases evolve. While the signal was very clear for the simulated data set S1 with a stationary signal (Fig. \ref{fig:evolved_phase}, top left panel), it was also clearly detected for sets S2 and S3 for which the evolution in the signal phase was 0.1 and 0.2 rad/week, respectively. With more rapidly evolving phase of 0.3 rad/week, the signal and its alias became less and less unique (S4), whereas even more rapidly evolving signals could not be detected as unique solutions despite the fact that the posterior densities showed hints of periodicities close to the period of the injected signal and its daily alias (Fig. \ref{fig:evolved_phase}, bottom panels).

The stationary signal (S1) corresponds to the only simulated case where the signal was detected as clearly as in the actual HARPS and HIRES radial velocity data. For the simulated set S1, the signal was detected with a logarithm of Bayes factor of 58.77 -- close to the value obtained for the actual data of 52.25. For sets S2 and S3,
this value decreased to 21.24 and 6.51 -- the latter one is only barely above the detection threshold of 5.01. Together with the fact that the signals loose their uniqueness when the signal varies more than 0.2 rad/week this implies that an evolving signal would be unlikely to cause the observed radial velocity variability. This suggests that only signals that do not vary as a function of time can be detected as strong probability maxima in the AD Leo radial velocity data (Fig. \ref{fig:posterior_period}).

Because linearly evolving phase is an unrealistic description of the potentially very complex patterns of starspot evolution, we also tested a simple stochastic variability model. When assuming that the phase of the signal was a random variable drawn from a Gaussian probability distribution centered at the current phase and a standard deviation $\sigma_{\psi}$ ranging from 0 to 0.7 rad as was the case for linearly evolving $\psi$ above. We updated the phase randomly after 2-4 rotation periods and again searched for signals in the generated artificial data sets. For such stochastic variability, the signal could not be detected in the data for $\sigma_{\psi} > 0.3$ rad. Although detection was possible for phase changing more slowly than that, our test again suggested that variable signals generally cannot be detected unless the rate of variability is low.

It thus appears that, unlike the photometric signal (Fig. \ref{fig:most}) that varies in phase, amplitude, and period, the radial velocity signal of AD Leo appears to be caused by a stationary process. At least, it can be said that the signal in AD Leo radial velocities is consistent with a stationary signal.

\section{AD Leo in the context of starspot observations and models}

We investigated the possibility of obtaining the observed photometric variability on AD Leo by modelling spots using the \emph{Doppler Tomography of Stars} (DoTS) program \citep{colliercameron2001}, which models both spectroscopic and photometric data. We used the model spectra of \citet{baraffe2015} to obtain spectral contrast-ratios for spots with 2900 and 3000 K and photosphere with 3400 K. In other words, for AD Leo, we assumed spots that are 400 K and 500 K cooler than the photosphere as indicated by \citet{berdyugina2005} and \citet{barnes2017}.

\begin{figure}
\center
\includegraphics[angle=270, width=0.49\textwidth, clip]{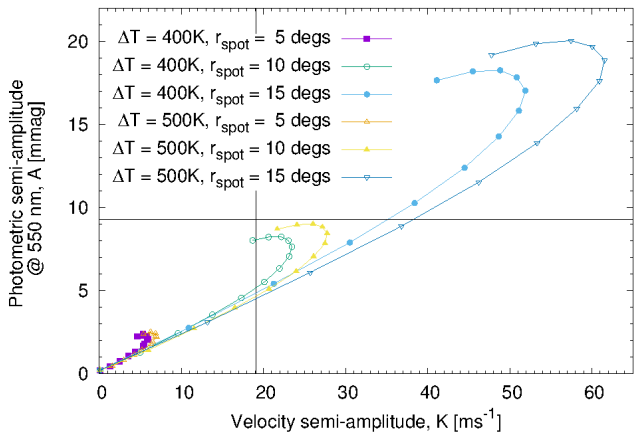}
\caption{Photometric amplitude as a function of radial velocity amplitude for a star with axial inclination, $i = 15^{\circ}$. Curves are plotted for spot radii of $r_{\rm spot} =$ 5, 10, and 15$^{\circ}$, with spot contrasts of
$T_{\rm phot} - T_{\rm spot}$ = $\Delta T$ = 400 and 500 K. The points defining each curve are for different spot latitudes, ranging from 30 to 90$^{\circ}$ (at 0,0) in steps of 5$^{\circ}$. The observed photometric and spectroscopic amplitudes are denoted by the horizontal and vertical lines, respectively.}\label{fig:spot_results}
\end{figure}

We tested scenarios for spots with radii of 5, 10, and 15$^{\circ}$, and fixed the stellar axial inclination at 15$^{\circ}$. The results are shown in Fig. \ref{fig:spot_results}, which shows the photometric amplitude ($A$) as a function of the spectroscopic radial velocity amplitude ($K$) induced by the spots. The points on each curve denote the amplitudes for a spot at latitude 90$^{\circ}$ (0,0) and then at successively lower latitudes down to 30$^{\circ}$ in 5$^{\circ}$ intervals. For spots with latitude $< 50^{\circ}$, the velocity amplitude begins to decrease while the photometric amplitude decreases slightly for spots at latitude $< 40^{\circ}$. This is a consequence of the centre-to-limb brightness variation. For a spot radius just larger than 10$^{\circ}$ and for $T_{\rm phot} - T_{\rm spot}$ = 500 K, a spot at latitude 40$^{\circ}$ is required to approximately reproduce the observed amplitude seen in the ASAS-N data and by \citet{spiesman1986}.

Only low-latitude spots are able to reproduce both the observed photometric and spectroscopic amplitudes for a star with $i = 15^{\circ}$. For the photometric amplitude observed in ASAS-N, larger spots at high latitudes (spot radius of 15$^{\circ}$ in Fig. \ref{fig:spot_results}) induce an RV variation that is much larger than observed. If the photometric amplitude is indeed underestimated (as suggested in Section \ref{sec:most}) and closer to $12$ mas as reported by \citet{spiesman1986} or more, it becomes even more difficult to explain the radial velocity and photometric signals with starspots. However, the radial velocity observations are not contemporaneous with the photometric observations. If the measured radial velocity amplitude of 19 ms$^{-1}$ is due to the reflex motion of a planet, our simulations suggest that when the high-cadence radial velocity observations were taken, the spot-induced photometric rotation amplitude must have been small. This fits well with our findings in Section \ref{sec:asas} where the stellar rotation is not always detected in the photometry implying a significantly smaller starspot.

Realistically, a single spot is probably an over-simplification. \citet{barnes2015} and \citet{barnes2017} show that rapidly rotating M dwarfs exhibit much smaller spots distributed at various longitudes and latitudes. It thus seems likely that photometric variability on AD Leo is only detectable when enough spots are present, i.e. when spot-coverage is increased and the star is thus at its brightness minimum (see Fig. \ref{fig:asas_data}). As demonstrated by MOST photometry (Section \ref{sec:most}) and the images in \citet{barnes2017}, spots can be stable on time-scales of a few days. Beyond that time-scale, the stability of individual spots or spot groups is rather poorly constrained. The \citet{barnes2017} observations for GJ791.2A (M4.5) separated by a year demonstrate a significant change in the distribution of spots at low and intermediate latitudes and that polar and circumpolar spot structures are changing significantly. Although these are rather limited observational constraints, these results do not appear to support the interpretation that long lived spots could exist for long enough to induce a stable RV signal.

\section{Comparison to other M dwarfs}

\subsection{Rapid rotators}\label{sec:reference_targets}

\begin{figure}
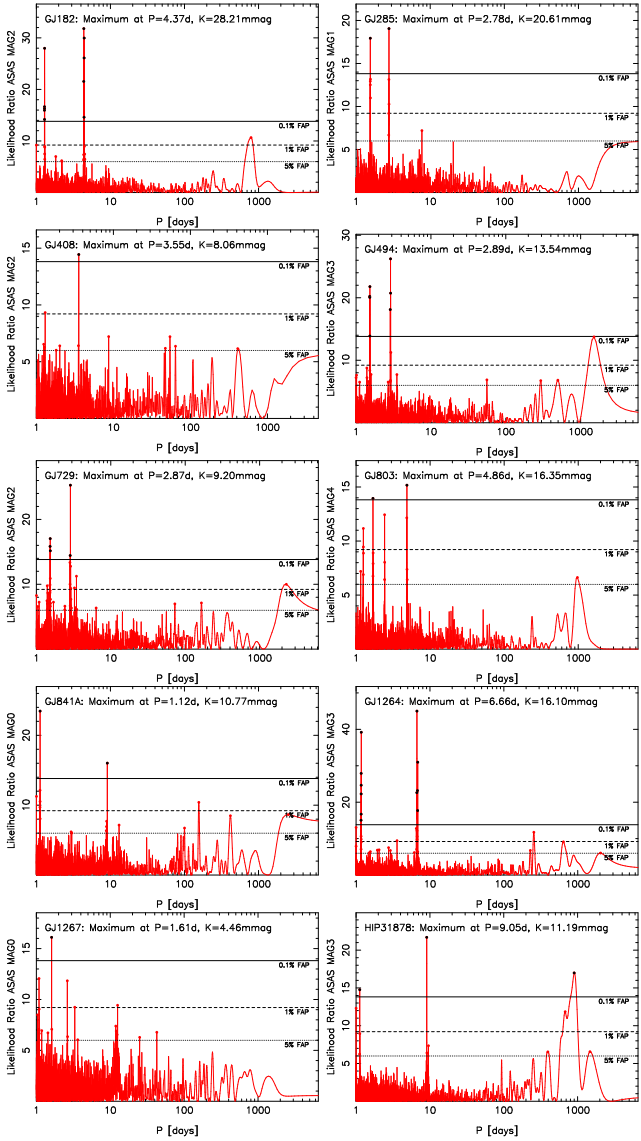

\center
\includegraphics[angle=270, width=0.23\textwidth]{GJ182_ASAS_mag2_mlresidual_wperiodog_logp.ps}
\includegraphics[angle=270, width=0.23\textwidth]{GJ285_ASAS_mag1_mlwperiodog_logp.ps}

\includegraphics[angle=270, width=0.23\textwidth]{GJ408_ASAS_mag2_mlwperiodog_logp.ps}
\includegraphics[angle=270, width=0.23\textwidth]{GJ494_ASAS_mag3_mlwperiodog_logp.ps}

\includegraphics[angle=270, width=0.23\textwidth]{GJ729_ASAS_mag2_mlresidual_wperiodog_logp.ps}
\includegraphics[angle=270, width=0.23\textwidth]{GJ803_ASAS_mag4_mlresidual_wperiodog_logp.ps}

\includegraphics[angle=270, width=0.23\textwidth]{GJ841A_ASAS_mag0_mlresidual_wperiodog_logp.ps}
\includegraphics[angle=270, width=0.23\textwidth]{GJ1264_ASAS_mag3_mlwperiodog_logp.ps}

\includegraphics[angle=270, width=0.23\textwidth]{GJ1267_ASAS_mag0_mlwperiodog_logp.ps}
\includegraphics[angle=270, width=0.23\textwidth]{HIP31878_ASAS_mag3_mlresidual_wperiodog_logp.ps}
\caption{Logarithm of the likelihood-ratio periodogram of the grade A ASAS-S photometry, with 5-$\sigma$ outliers removed for selected nearby M dwarfs for which there is evidence for a photometric rotation period with $P_{\rm rot} < 10$ days.}\label{fig:photometric_rotation}
\end{figure}

The ASAS survey has been used to identify several nearby M dwarfs with short ($<$10 days) photometric rotation periods \citep{kiraga2007}. We identified 10 targets in a sample of 360 nearby M dwarfs \citep{tuomi2017} for which the ASAS-S photometry shows evidence for periodicities shorter than 10 days (Table \ref{tab:phot_periods}). We have tabulated the corresponding significant (exceeding 0.1\% FAP) photometric periodicities with $P_{\rm rot} < 10$ days in Table \ref{tab:phot_periods} and interpret them as photometric rotation periods of the stars with period $P_{\rm rot}$. We have also plotted the likelihood-ratio periodograms of these targets in Fig. \ref{fig:photometric_rotation} to visually demonstrate the significances of the detected photometric rotation periods. All these targets have also been observed spectroscopically with HARPS, HIRES, and/or the \emph{Planet Finding Spectrograph} (PFS), and we thus examined the radial velocity data sets in order to search for counterparts of the photometric periodicities.

\begin{table}
\begin{minipage}{0.45\textwidth}
\begin{center}
\caption{List of M dwarfs included in the HARPS, HIRES, and/or PFS radial velocity surveys for which photometric rotation periods with $P < 10$ days are known based on ASAS-S photometry. $A$ denotes the amplitude of the photometric signal. $N$, and $\sigma(v)$ denote the number of radial velocity data available for the target and the standard deviation of the velocities, respectively.}\label{tab:phot_periods}
\begin{tabular}{lccccc}
\hline \hline
Target & $P_{\rm rot}$ & $A$ & $N$ & $\sigma(v)$ \\
& (d) & (mmag) &  & (ms$^{-1}$) \\
\hline
GJ 182 & 4.37 & 28.21 & 9 & 217.1 \\
GJ 285 & 2.78 & 20.61 & 30 & 84.7 \\
GJ 408 & 3.55 & 8.06 & 68 & 5.5 \\
GJ 494 & 2.89 & 13.54 & 2 & -- \\
GJ 729 & 2.87 & 9.20 & 97 & 23.9 \\
GJ 803 & 4.86 & 16.35 & 35 & 136.6 \\
GJ 841A & 1.12 & 10.77 & 6 & -- & \\
GJ 1264 & 6.66 & 16.10 & 1 & -- & \\
GJ 1267 & 1.61 & 4.46 & 39 & 5.0 \\
HIP 31878 & 9.15 & 11.19 & 26 & 40.1 \\
\hline \hline
\end{tabular}
\end{center}
\end{minipage}
\end{table}

\begin{figure}
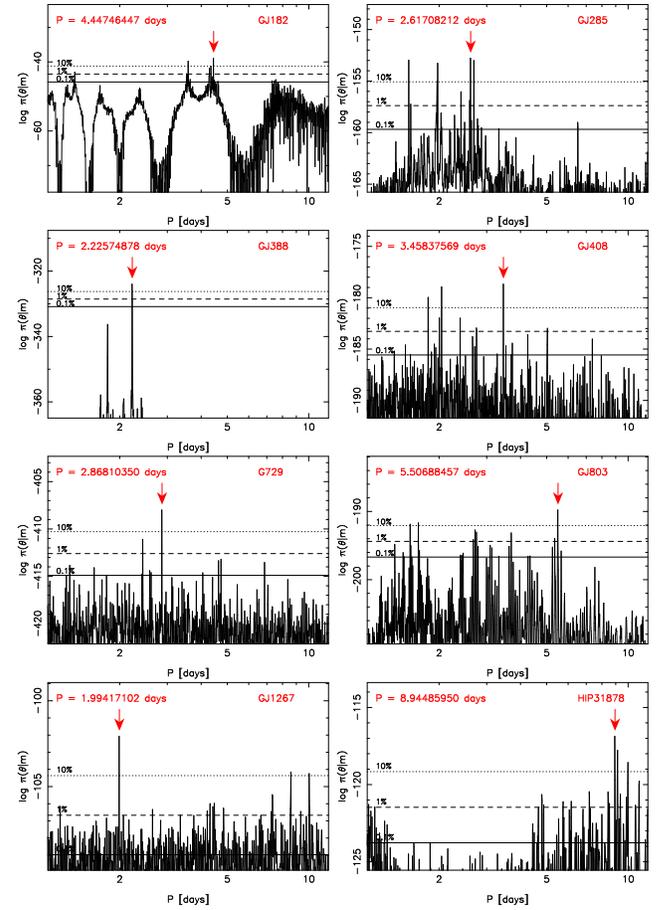

\center
\includegraphics[angle=270, width=0.23\textwidth]{rv_GJ182_01_pcurve_b.ps}
\includegraphics[angle=270, width=0.23\textwidth]{rv_GJ285_01_pcurve_b.ps}

\includegraphics[angle=270, width=0.23\textwidth]{rv_GJ388_01_pcurve_b.ps}
\includegraphics[angle=270, width=0.23\textwidth]{rv_GJ408_01_pcurve_b.ps}

\includegraphics[angle=270, width=0.23\textwidth]{rv_GJ729_01_pcurve_b.ps}
\includegraphics[angle=270, width=0.23\textwidth]{rv_GJ803_01_pcurve_b.ps}

\includegraphics[angle=270, width=0.23\textwidth]{rv_GJ1267_01_pcurve_b.ps}
\includegraphics[angle=270, width=0.23\textwidth]{rv_HIP31878_01_pcurve_b.ps}
\caption{Estimated posterior probability density as a function of the period of the signal. The horizontal lines denote the 10\%, (dotted), 1\% (dashed), and 0.1\% (solid) equiprobability thresholds with respect to the maximum denoted by using a red arrow.}\label{fig:posterior_targets}
\end{figure}

We also investigated whether the available radial velocities of the targets in Table \ref{tab:phot_periods} showed evidence for signals that could be interpreted as counterparts of the photometric rotation periods. Apart from GJ 494 and GJ 1264 that only had 2 and 1 radial velocity measurements available, respectively, and GJ 841A whose HARPS spectra were contaminated and resulted in radial velocities varying at a 10 kms$^{-1}$ level, we searched for such counterparts of photometric rotation periods with posterior samplings. The resulting estimated posterior densities as functions of signal periods are shown in Fig. \ref{fig:posterior_targets}. We also show the posterior given AD Leo (GJ 388) data for the same period space between 1 and 12 days.

It can be seen in Fig. \ref{fig:posterior_targets} that only AD Leo (GJ 388; left column, second panel from the top) shows evidence for a unique radial velocity signal in the sense that there are no local maxima exceeding the 0.1\% equiprobability threshold of the global maximum. However, there seems to be a reasonable correspondence in the sense that all targets have global posterior maxima at or near the detected photometric periodicities even though the maxima in the radial velocities are far from unique (Fig. \ref{fig:posterior_targets}). It is thus likely that the radial velocities contain periodic variability corresponding to the photometric rotation periods. But AD Leo is the exception in this sense as it is the only one with a unique short-period radial velocity signal.

We interpret this as an indication that such rapidly rotating M dwarfs do not readily produce clearly distinguishable radial velocity signals at or near the rotation period. This is the case even when the photometric rotation period is readily detectable with ASAS photometry. When searching for signals in the data of the comparison stars (Fig. \ref{fig:posterior_targets}) we obtain broadly similar results as for our artificial data sets with evolving signals (Fig. \ref{fig:evolved_phase}), i.e. there are no unique and significant probability maxima. AD Leo clearly represents an exception in both these respects.

It was not possible to study the dependence of the comparison target radial velocity signals on spectral wavelength as was done for AD Leo (Fig. \ref{fig:wavelength_dependence}). We calculated the radial velocities for the six wavelength intervals, each corresponding to twelve HARPS orders, for GJ 803 that had the most HARPS observations available ($N = 20$ after removing outliers). Significant periodicities could not be detected in any of these six sets and we thus could not study the wavelength dependence of signals caused by purely stellar rotation. However, the probability maximum corresponded to different periods in each of the six wavelength intervals. These periods ranged from 2.8 to 6.3 days but it remains uncertain whether any of them actually correspond to the photometric rotation period of the star of 4.86 days.

\subsection{The slowly rotating planet host GJ 674}\label{sec:GJ674}

Finally, we analysed the HARPS radial velocity data of GJ 674, a quiescent slowly rotating M dwarf, that has been reported to be a host to a candidate planet with a minimum mass of 11.09 M$_{\oplus}$ with an orbital period of 4.6938$\pm$0.007 days \citep{bonfils2007}. Moreover, as discussed by \citet{bonfils2007} and \citet{mascareno2015}, the HARPS radial velocities of GJ 674 also show evidence for the star's rotation period. This period was estimated to be 34.8467$\pm$0.0324 days by \citet{bonfils2007} but other authors provide slightly different estimates of 32.9$\pm$0.1 days \citep{mascareno2015} based on spectroscopic activity indices and 33.29 days \citep{kiraga2007} based on ASAS photometry. Because this target provides an ideal test-case for studying the differences between a planet- and rotation-induced signals, we examined the wavelength-dependence of the two signals based on radial velocity data calculated for different wavelength ranges as we did for AD Leo in Section \ref{sec:colour-time}.

As is expected for a planetary signal, the signal of GJ 674 b was consistently detected in all but the bluest twelve HARPS orders. For the five reddest sets of twelve orders, we obtained a consistent periodicity of 4.6950$\pm$0.0002 days and an amplitude of 8.7$\pm$0.3 ms$^{-1}$ that also agrees very well with the solution reported by \citet{bonfils2007}.

\begin{figure}
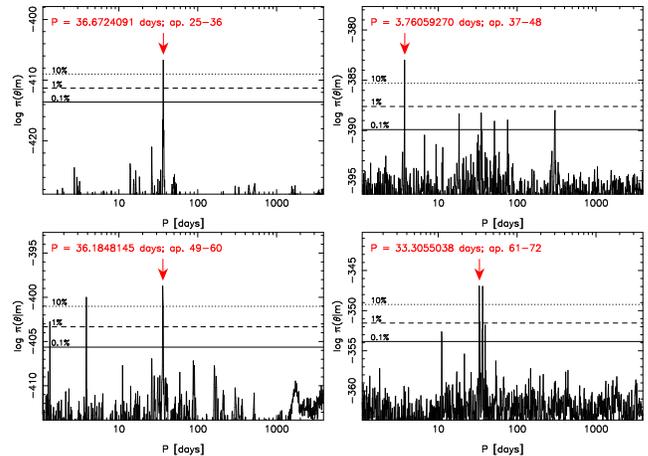

\center
\includegraphics[angle=270, width=0.23\textwidth]{rv_GJ674_6AP3_02_pcurve_c.ps}
\includegraphics[angle=270, width=0.23\textwidth]{rv_GJ674_6AP4_03_pcurve_d.ps}

\includegraphics[angle=270, width=0.23\textwidth]{rv_GJ674_6AP5_02_pcurve_c.ps}
\includegraphics[angle=270, width=0.23\textwidth]{rv_GJ674_6AP6_03_pcurve_d.ps}
\caption{Estimated posterior probability densities used to determine the global maxima corresponding to the rotation period in the HARPS radial velocities of GJ 674. The signal of GJ 674 b has been accounted for and the red arrows indicate the positions of the most prominent secondary maxima. The aperture ranges correspond to the four redmost sets of twelve HARPS orders.}\label{fig:GJ674_signals}
\end{figure}

However, the rotation-induced radial velocity signal of GJ 674 was found to be dependent on wavelength. The rotation signal could not be detected in the two bluest sets of twelve orders, 1-12 and 13-24, at all, although the latter showed hints of a periodicity of 36.67 days corresponding to the highest probability maximum in the period space. In radial velocities calculated for orders 25-36, 37-48, 49-60, and 61-72, we detected periodicities of 36.66$\pm$0.19, 3.6708$\pm$0.0004, 36.18$\pm$0.05, and 33.33$\pm$0.03, respectively, although the last set of orders had two roughly equally high probability maxima with an alternative solution at a period of 36.66$\pm$0.03 days (Fig. \ref{fig:GJ674_signals}). According to these results, the rotation-induced signal is not found consistently at the same period but varies as a function of wavelength. Moreover, a signal corresponding to the star's rotation period could not be identified in the radial velocities calculated for orders 37-48 but another signal near four days was present instead. This signal was also present in orders 49-60 as a secondary solution (Fig. \ref{fig:GJ674_signals}, bottom left panel). We did not observe significant differences in the amplitudes of the signals as a function of wavelength because the amplitude of the rotation-induced signal had an amplitude below 3 ms$^{-1}$ with uncertainties of roughly 1 ms$^{-1}$, making the available precision insufficient for determining possible differences in amplitude as a function of wavelength. These results demonstrate that stellar rotation does not readily produce wavelength-invariant radial velocity signals and that the signal observed in AD Leo radial velocities is thus likely caused by a planet orbiting the star.

\begin{figure}
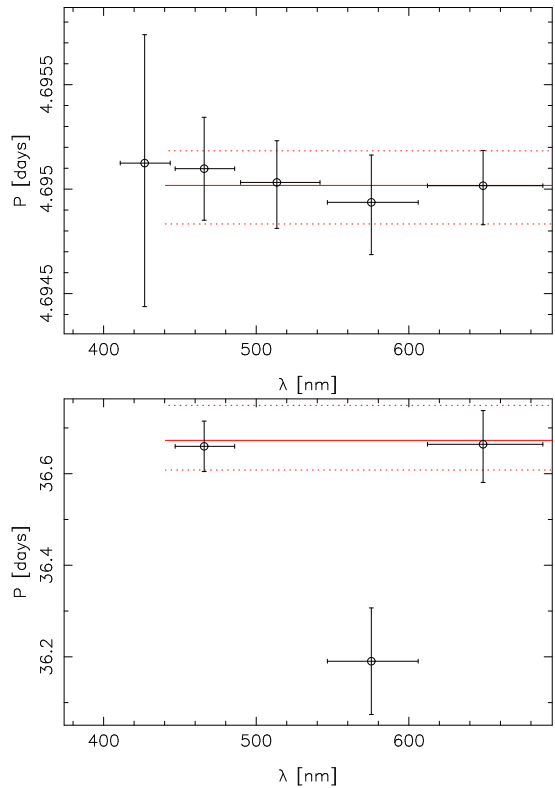

\center
\includegraphics[angle=270, width=0.40\textwidth]{GJ674_period.ps}

\includegraphics[angle=270, width=0.40\textwidth]{GJ674_period2.ps}
\caption{As in Fig. \ref{fig:wavelength_dependence} but for the period parameters of GJ 674 signals corresponding to planet candidate GJ 674b (top panel) and the rotation-induced signal (bottom panel).}\label{fig:GJ674_period}
\end{figure}

We further highlight the wavelength dependence and independence of the Keplerian signal in GJ 674 data and the rotation-induced signal near 37 days in Fig. \ref{fig:GJ674_period}. We note that the estimated second period is not shown for orders 37-48 in bottom panel of Fig. \ref{fig:GJ674_period} because there are no significant signals at or near the stellar rotation period and the corresponding second signal for these orders has a period of 3.67 days.

\section{Discussion}

We have presented analyses of ASAS V-band photometry, MOST photometry, and HARPS and HIRES radial velocities of AD Leo. Although AD Leo is a rapidly rotating star with a rotation period of approximately 2.23 days \citep[see][and the current work]{spiesman1986,morin2008,hunt2012,bonfils2013,reiners2013}, we only see evidence for a photometric rotation period of the star in ASAS-N and MOST photometry (Fig. \ref{fig:asas_periodogram} when the star is at a brightness maximum of its activity cycle. This might be due to the brightness maximum corresponding to a lower starspot coverage, making the photometric rotation period more visible in the data. Non-detection of this signal in the ASAS-S photometry might indicate that there are too many spots on the stellar surface, or that they are too variable, during the brightness minimum to determine the photometric rotation period.

The radial velocities of AD Leo contain a unique and highly significant signal, that appears to be time- and wavelength-invariant, at a period of 2.22567 [2.22556, 2.22593] days (Fig. \ref{fig:posterior_period}) coinciding with the rotation period of the star. Our results indicate that the radial velocity signal of AD Leo is independent of spectral wavelength range (Fig. \ref{fig:wavelength_dependence}) and also time-invariant, making it unlikely to have been caused by stellar activity and starspots co-rotating on the stellar surface. Our tests with simulated data indicate that only stationary periodic processes can give rise to such a clear radial velocity signal as we observed in the HARPS and HIRES radial velocities (Fig. \ref{fig:evolved_phase}).

We consider it difficult to interpret stellar rotation as the origin of the time- and wavelength-invariant radial velocity signal. The photometric signal varies much more on short time-scales (Section \ref{sec:most}) than the the radial velocity signal appears to do over a baseline of thousands of days (Section \ref{sec:colour-time}). Our modelling also demonstrates that the observed amplitudes of signals in radial velocity and photometry data are difficult to explain by starspots (Fig \ref{fig:spot_results}). Instead, we propose an alternative hypothesis: the radial velocity signal of AD Leo is caused by a planet orbiting the star, locked in spin-orbit resonance.

Typically such coincidences have been interpreted by simply stating that the radial velocity signal is caused by stellar rotation. However, spin-orbit synchronisation might lead to scenarios where such coincidences occur as well \citep{mcquillan2013,walkowicz2013}. According to \citet{walkowicz2013}, who identified several potential cases in \emph{Kepler} data where stellar rotation periods are equal to or twice the planetary orbital periods, spin-orbit synchronisation only happens for planet candidates with radii $R > 6$R$_{\oplus}$ implying that such planets would be closer in radius to Saturn than Neptune. This further suggests that, should AD Leo b exist, it is probably larger than Neptune in radius and thus also more massive than the minimum mass of 19.7$\pm$1.9 M$_{\oplus}$ as also suggested by the fact that AD Leo is oriented almost pole-on. Given no spin-orbit misalignment and an inclination of $15.5 \pm 2.5^{\circ}$ this implies a true mass of 75.4$\pm$14.7 M$_{\oplus}$ or 0.237$\pm$0.047 M$_{\rm Jup}$. 

Such hot giant planets orbiting M dwarfs, although rare with occurrence rate $<$ 1\% planets per star \citep{dressing2015}, have also been confirmed transiting \emph{Kepler} targets Kepler-45 and Kepler-785 \citep{johnson2012,morton2016}. However, the occurrence rate of short-period giant planets around young, active M dwarfs is not known because the corresponding targets would mostly be selected against when choosing radial velocity targets. The same is the case with \emph{Kepler} transit photometry, for which the automatic data reduction pipelines would likely reject transit signals in spin-orbit resonance cases because of their interpretation as astrophysical false positives (binary stars).

When looking at radial velocities of other similar nearby M dwarfs that are known to be rapid rotators based on detections of photometric rotation periods in ASAS-S V-band photometry, such rotation periods do not generally give rise to unique and significant radial velocity signals (Fig. \ref{fig:posterior_targets}). AD Leo does not seem to fit into this pattern because the comparison stars that have stable photometric signatures of rotation periods in ASAS data do not show strong evidence for unique periodic signals in their respective radial velocities (Fig. \ref{fig:posterior_targets}). Moreover, the rotation-induced radial velocity signal of the slow rotator GJ 674 is not independent of wavelength, indicating that stellar rotation cannot be expected to produce wavelength-invariant radial velocity signals. These results suggest that the radial velocity signal of AD Leo is probably caused, at least partially, by a planet rather than being exclusively a consequence of stellar rotation.

We considered the possibility that the nearly pole-on orientation makes AD Leo different from the reference targets in Section \ref{sec:reference_targets}. However, this scenario is unlikely because the amplitude of a starspot-induced radial velocity signal is proportional to $\sin i$ and thus decreases to zero as inclination of the rotation axis approaches zero. It should therefore be expected that stars that are further from pole-on orientation, i.e. the reference targets for which this is the case on average (assuming random orientation in space), show greater, more easily detectable, periodic radial velocity variability caused by co-rotation of starspots on the stellar surface. Yet, the exact opposite is observed.

We consider it unlikely that the unique and time- and wavelength-invariant radial velocity signal of AD Leo at a period of 2.23 days could be caused by stellar rotation. Rather, it seems probable that it is in fact caused by a planet with a mass of 0.237$\pm$0.047 M$_{\rm Jup}$ orbiting the star. If this interpretation is correct, AD Leo, due to its vicinity with $d = 4.9$ pc, is an important benchmark target for studying spin-orbit resonances and star-planet interactions in early stages of stellar evolution. Our results for GJ 674 also suggest that rotation-induced radial velocity signals can be differentiated from Keplerian ones by looking at the dependence of the signals on spectral wavelength.

\acknowledgements
MT and HRAJ are supported by grants from the Leverhulme Trust (RPG-2014-281) and the Science and Technology Facilities Council (ST/M001008/1). SSV gratefully acknowledges support from NSF grants AST-0307493 and AST-0908870. The work herein is partially based on observations obtained at the W. M. Keck Observatory, which is operated jointly by the University of California and the California Institute of Technology, and we thank the UC-Keck and NASA-Keck Time Assignment Committees for their support. This research has made use of the Keck Observatory Archive (KOA), which is operated by the W. M. Keck Observatory and the NASA Exoplanet Science Institute (NExScI), under contract with the National Aeronautics and Space Administration. We would like to thank Nicholas M. Hunt-Walker for providing the MOST data. We also wish to extend our special thanks to those of Hawaiian ancestry on whose sacred mountain of Mauna Kea we are privileged to be guests. Without their generous hospitality, the Keck observations presented herein would not have been possible. This research has made use of the SIMBAD database, operated at CDS, Strasbourg, France. We are very grateful to the unanimous reviewer who presented suggestions and constructive criticism greatly improving the paper.

\begin{table*}
\caption{HARPS data of AD Leo.}\label{tab:HARPS_GJ388}
\begin{minipage}{\textwidth}
\begin{center}
\begin{tabular}{lccccc}
\hline \hline
$t$ [JD-2450000] & $v$ [ms$^{-1}$] & $\sigma_{v}$ [ms$^{-1}$] & BIS [ms$^{-1}$] & FWHM [ms$^{-1}$] & S [ms$^{-1}$] \\
\hline
2452986.8579 &  27.05 & 1.86 & -15.03 & 3521.66 &  8.208 \\
2453511.5468 & -15.67 & 1.15 & -10.57 & 3527.78 &  8.891 \\
2453520.5205 & -13.86 & 1.06 &  -6.66 & 3524.64 &  8.232 \\
2453543.4808 &  15.27 & 1.14 &  -6.62 & 3530.43 &  6.919 \\
2453544.4518 &  -6.06 & 0.86 &  -6.23 & 3534.29 &  7.811 \\
2453550.4596 &  41.73 & 2.07 & -14.23 & 3522.10 &  7.026 \\
2453728.8648 &  38.93 & 1.36 & -10.10 & 3525.78 & 10.308 \\
2453758.7540 & -18.71 & 0.88 &  -8.21 & 3523.30 &  7.151 \\
2453760.7547 &  -9.90 & 0.81 &  -6.21 & 3533.71 &  8.565 \\
2453761.7800 &  19.20 & 0.83 & -10.04 & 3520.08 &  7.369 \\
2453783.7255 &  -4.53 & 0.71 &  -6.76 & 3524.50 &  7.576 \\
2453785.7264 & -15.76 & 0.86 &  -9.08 & 3529.40 &  6.955 \\
2453809.6599 &   0.00 & 0.52 &  -9.49 & 3525.41 &  7.037 \\
2453810.6767 &  13.63 & 1.26 &  -8.86 & 3514.12 & 11.415 \\
2453811.6750 &   9.90 & 0.62 & -12.42 & 3529.17 &  8.512 \\
2453812.6637 &  -5.84 & 0.71 &  -9.97 & 3514.26 &  8.135 \\
2453813.6584 &  22.38 & 0.73 &  -5.76 & 3532.84 &  7.916 \\
2453814.6540 & -16.21 & 1.06 &  -6.89 & 3525.54 &  6.941 \\
2453815.5702 &  26.82 & 0.97 & -16.25 & 3522.02 &  7.188 \\
2453815.6211 &  26.18 & 0.85 & -11.89 & 3527.75 &  7.324 \\
2453815.7354 &  25.17 & 0.82 & -10.47 & 3516.77 &  7.135 \\
2453816.5433 & -11.46 & 0.62 &  -9.59 & 3521.54 &  7.795 \\
2453816.6552 & -16.34 & 0.71 &  -6.39 & 3522.61 &  7.150 \\
2453816.7211 & -17.59 & 0.78 &  -6.83 & 3519.82 &  7.263 \\
2453817.5500 &  24.85 & 0.77 &  -8.59 & 3523.99 &  7.423 \\
2453817.6749 &  27.92 & 0.66 & -10.12 & 3525.81 &  8.399 \\
2453829.6146 &   0.88 & 0.77 &  -9.81 & 3520.37 &  7.688 \\
2453830.5397 &  -4.17 & 1.14 &  -9.00 & 3504.41 &  9.871 \\
2453831.6694 &  10.38 & 0.55 & -12.54 & 3521.13 &  6.977 \\
2453832.6497 &  -8.13 & 0.69 &  -5.59 & 3511.21 &  7.954 \\
2453833.6269 &  24.45 & 0.72 & -11.10 & 3523.54 &  7.195 \\
2453834.6103 & -23.28 & 1.33 &  -4.47 & 3517.78 &  9.454 \\
2453835.6551 &  28.84 & 0.91 & -12.59 & 3518.32 &  7.368 \\
2453836.6165 & -15.98 & 0.65 &  -3.33 & 3524.45 &  6.811 \\
2453861.5934 &   1.93 & 0.70 &  -9.12 & 3502.93 &  7.065 \\
2453863.5687 & -17.61 & 0.93 &  -6.54 & 3519.25 &  8.321 \\
2453864.5340 &  30.04 & 0.93 & -13.78 & 3517.17 &  7.522 \\
2453867.5417 & -10.17 & 0.76 &  -7.56 & 3514.19 &  7.097 \\
2453868.5177 &  22.94 & 0.71 & -12.73 & 3517.50 &  7.330 \\
2453871.5622 &  19.51 & 0.62 &  -8.47 & 3517.11 &  7.414 \\
2456656.8493 & -31.74 & 1.16 &  -0.98 & 3542.60 &  7.184 \\
2456656.8602 & -28.41 & 1.03 &  -3.16 & 3539.87 &  7.314 \\
2456657.8522 &  40.05 & 1.28 & -18.92 & 3544.42 &  8.095 \\
2456658.8644 & -38.27 & 0.87 &  -0.27 & 3547.32 &  7.422 \\
2456658.8755 & -38.61 & 1.52 &   0.29 & 3547.48 &  7.426 \\
2456659.8581 &  42.94 & 1.58 & -15.30 & 3539.94 &  7.658 \\
2456797.5118 & -24.02 & 0.74 &  -1.73 & 3541.87 &  8.375 \\
\hline \hline
\end{tabular}
\end{center}
\end{minipage}
\end{table*}

\begin{table*}
\caption{Example of HARPS radial velocity of AD Leo for all 72 individual orders $v_{i}, i=1, ..., 72$.}\label{tab:HARPS_GJ388_orders}
\begin{minipage}{\textwidth}
\begin{center}
\begin{tabular}{lccccccc}
\hline \hline
$t$ [JD-2450000] & $v_{1}$ [ms$^{-1}$] & $\sigma_{v_{1}}$ [ms$^{-1}$] & $v_{2}$ [ms$^{-1}$] & $\sigma_{v_{2}}$ [ms$^{-1}$] & $v_{3}$ [ms$^{-1}$] & $\sigma_{v_{3}}$ [ms$^{-1}$] & \\
\hline
2986.8579 &  502.80 & 300.23 & -1153.97 & 276.03 &  190.56 & 289.89 & ... \\
3511.5468 &  -38.00 & 141.87 &   -94.04 & 133.81 &  521.06 & 126.51 & ... \\
3520.5205 &  158.29 & 191.61 &   471.33 & 187.46 &  278.08 & 169.40 & ... \\
3543.4808 &  138.40 & 240.76 &    17.90 & 227.65 &  -75.73 & 216.12 & ... \\
3544.4518 & -377.68 & 171.55 &  -354.19 & 162.54 & -156.08 & 148.84 & ... \\
3550.4596 & 1176.80 & 321.81 &  -744.10 & 286.01 &  369.98 & 296.29 & ... \\
3728.8648 &   -8.87 & 162.99 &   195.14 & 155.38 &  -16.01 & 144.67 & ... \\
3758.7540 &   81.83 & 129.82 &    50.39 & 127.00 &  -21.92 & 111.59 & ... \\
3760.7547 &  -18.18 & 120.27 &  -472.43 & 116.04 &  168.59 & 107.86 & ... \\
3761.7800 &  184.64 & 107.38 &    19.01 & 103.78 & -113.55 &  90.69 & ... \\
3783.7255 & -158.69 & 113.05 &   103.10 & 112.32 &  150.24 &  96.51 & ... \\
3785.7264 &   -1.90 & 111.54 &  -155.30 & 115.46 &  257.99 &  99.78 & ... \\
3809.6599 &   49.90 &  79.20 &  -143.19 &  82.32 &   24.83 &  69.91 & ... \\
3810.6767 &  314.57 & 108.86 &   296.13 &  96.35 &  110.75 &  84.68 & ... \\
3811.6750 &  123.42 &  98.04 &   200.36 &  93.21 & -131.53 &  81.88 & ... \\
3812.6637 & -157.04 & 166.80 &   538.03 & 159.24 & -261.14 & 145.15 & ... \\
3813.6584 &  231.57 &  89.82 &   -45.62 &  88.21 &  -17.41 &  78.00 & ... \\
3814.6540 &  -91.03 & 152.77 &  -217.45 & 148.51 &  -72.68 & 132.42 & ... \\
3815.5702 &  -14.76 & 121.71 &   232.38 & 119.64 & -153.38 & 107.72 & ... \\
3815.6211 & -217.01 & 103.38 &   -41.35 & 100.24 & -227.38 &  88.13 & ... \\
3815.7354 & -400.83 & 109.28 &     5.05 & 105.98 &  -46.12 &  90.78 & ... \\
3816.5433 &   35.14 & 127.58 &   194.62 & 123.77 & -150.30 & 111.89 & ... \\
3816.6552 & -186.00 & 107.40 &  -201.14 & 109.34 &  -39.90 &  92.84 & ... \\
3816.7211 &  127.14 & 116.73 &    65.59 & 118.16 & -218.59 & 100.93 & ... \\
3817.5500 &  154.66 &  94.25 &   -95.49 &  95.87 &  -76.58 &  81.27 & ... \\
3817.6749 &   84.06 & 103.83 &    83.03 &  99.67 &  100.74 &  88.41 & ... \\
3829.6146 & -247.02 &  97.13 &    76.68 &  95.42 &  118.33 &  81.55 & ... \\
3830.5397 &  164.12 & 119.88 &   -63.25 & 112.66 & -100.52 & 101.58 & ... \\
3831.6694 &  224.15 &  89.65 &  -165.35 &  91.34 &  177.34 &  76.56 & ... \\
3832.6497 &  124.92 &  92.06 &  -128.90 &  95.22 & -233.24 &  82.47 & ... \\
3833.6269 & -130.12 &  79.78 &    25.88 &  76.59 &   65.00 &  66.56 & ... \\
3834.6103 & -208.36 &  85.12 &  -248.17 &  79.18 &   99.07 &  68.95 & ... \\
3835.6551 &  -73.31 &  98.09 &   -53.44 &  95.03 & -223.53 &  84.78 & ... \\
3836.6165 & -166.97 &  87.31 &   118.87 &  87.04 & -320.86 &  75.65 & ... \\
3861.5934 &  629.82 & 127.62 &   106.85 & 126.00 & -329.65 & 110.16 & ... \\
3863.5687 &  -70.57 & 103.04 &  -450.80 &  96.21 &   76.85 &  85.86 & ... \\
3864.5340 &   25.72 & 106.61 &   156.79 & 108.93 &  147.68 &  92.01 & ... \\
3867.5417 & -415.38 & 118.93 &   107.27 & 115.52 & -124.68 & 101.36 & ... \\
3868.5177 &  108.62 & 115.43 &  -265.25 & 112.28 & -256.27 &  98.14 & ... \\
3871.5622 &  -94.14 & 105.93 &   -71.85 & 104.76 &  118.22 &  92.04 & ... \\
6656.8493 &  291.80 & 204.95 &  -289.60 & 194.67 &  142.49 & 175.07 & ... \\
6656.8602 &  247.43 & 200.15 &  -233.94 & 189.16 &  542.26 & 168.08 & ... \\
6657.8522 & -208.53 & 212.18 &   534.53 & 194.33 &  278.32 & 173.81 & ... \\
6658.8644 & -313.55 & 204.48 &  -212.75 & 197.12 & -231.53 & 169.65 & ... \\
6658.8755 &  -42.68 & 197.72 &   117.97 & 189.36 & -112.47 & 162.47 & ... \\
6659.8581 & -465.41 & 291.66 &  1210.13 & 257.40 &   -2.09 & 247.89 & ... \\
6797.5118 &  126.76 & 105.52 &   271.71 & 102.69 &  216.04 &  90.55 & ... \\
\hline \hline
\end{tabular}
\end{center}
\end{minipage}
\end{table*}

\begin{table*}
\caption{Example of HARPS data of GJ 674.}\label{tab:HARPS_GJ674}
\begin{minipage}{\textwidth}
\begin{center}
\begin{tabular}{lccccc}
\hline \hline
$t$ [JD-2450000] & $v$ [ms$^{-1}$] & $\sigma_{v}$ [ms$^{-1}$] & BIS [ms$^{-1}$] & FWHM [ms$^{-1}$] & S [ms$^{-1}$] \\
\hline
2453158.7520 &   9.73 & 0.73 &  -8.99 & 3018.96 & 1.318 \\
2453205.5995 &   8.41 & 0.32 &  -6.72 & 3065.91 & 1.665 \\
2453237.5635 &   7.53 & 0.83 &  -9.55 & 3070.47 & 1.382 \\
2453520.7932 &   3.66 & 0.54 &  -8.49 & 3065.37 & 1.611 \\
2453580.5685 &   7.06 & 0.41 &  -8.13 & 3068.93 & 1.528 \\
2453813.8477 & -10.69 & 0.45 &  -8.55 & 3063.31 & 1.130 \\
2453816.8691 &  -1.00 & 0.49 &  -6.33 & 3059.31 & 1.086 \\
2453861.8085 &   4.73 & 0.47 & -10.11 & 3064.74 & 1.204 \\
2453862.7863 &  10.41 & 0.38 &  -7.77 & 3066.12 & 1.563 \\
2453863.8105 &   6.07 & 0.54 &  -8.92 & 3066.16 & 1.254 \\
2453864.7669 &  -8.34 & 0.47 &  -7.75 & 3069.48 & 1.432 \\
2453865.8114 &  -0.96 & 0.48 & -10.58 & 3070.48 & 1.353 \\
2453866.7558 &   4.98 & 0.37 &  -8.74 & 3071.30 & 1.755 \\
2453867.8481 &   6.56 & 0.44 &  -8.07 & 3069.35 & 1.461 \\
2453868.8393 &  -4.07 & 0.43 &  -5.99 & 3075.55 & 1.662 \\
2453869.8038 & -11.75 & 0.49 &  -8.93 & 3080.39 & 2.052 \\
2453870.7388 &  -3.83 & 0.65 &  -9.05 & 3079.69 & 1.651 \\
2453871.8606 &   1.96 & 0.52 &  -7.08 & 3076.56 & 1.379 \\
2453882.7445 &  -4.55 & 0.44 &  -7.59 & 3062.54 & 1.184 \\
2453886.7489 &   2.98 & 0.44 &  -8.69 & 3063.01 & 1.081 \\
2453887.7915 &  -7.14 & 0.37 &  -7.69 & 3060.73 & 1.006 \\
2453917.7229 &  -3.22 & 0.47 &  -7.40 & 3058.93 & 1.121 \\
2453919.7271 &   5.47 & 0.66 &  -4.99 & 3063.84 & 1.014 \\
2453921.6271 &  -8.82 & 0.39 &  -8.70 & 3059.50 & 1.091 \\
2453944.5860 & -13.05 & 0.66 &  -6.79 & 3072.41 & 1.529 \\
2453947.5906 &   8.07 & 1.34 &  -8.24 & 3076.75 & 1.304 \\
2453950.6474 &  -3.86 & 0.49 &  -8.61 & 3064.41 & 1.285 \\
2453976.5244 &  -2.45 & 0.46 &  -8.58 & 3074.71 & 1.410 \\
2453980.5691 &  -0.09 & 0.56 &  -8.55 & 3075.49 & 1.265 \\
2453981.5927 & -11.94 & 0.53 &  -6.96 & 3058.65 & 1.383 \\
2453982.6162 & -16.16 & 0.55 &  -7.27 & 3058.62 & 1.267 \\
2453983.5287 &  -5.36 & 0.66 &  -8.48 & 3061.14 & 1.068 \\
2454167.8783 &   2.75 & 0.45 &  -7.76 & 3073.49 & 1.336 \\
2454169.8879 & -13.41 & 0.44 &  -7.93 & 3072.81 & 1.410 \\
2454171.8966 &   1.79 & 0.46 &  -6.72 & 3069.83 & 1.271 \\
2454173.8674 &  -1.82 & 0.41 &  -8.59 & 3069.64 & 1.297 \\
2454340.6256 &  -0.18 & 0.39 &  -9.31 & 3070.89 & 1.330 \\
2454342.5927 &   3.43 & 0.52 &  -7.58 & 3070.21 & 1.353 \\
2454347.5335 &  -5.20 & 0.62 &  -7.29 & 3066.48 & 1.237 \\
2454349.6053 &  -9.33 & 0.51 &  -7.85 & 3055.05 & 1.150 \\
2454388.4992 &   3.41 & 0.61 &  -5.06 & 3057.18 & 1.579 \\
2454589.8983 &  -0.64 & 0.47 &  -7.35 & 3060.15 & 1.347 \\
2454666.7202 &   0.52 & 0.55 &  -7.10 & 3070.37 & 1.236 \\
2454732.4793 &   2.08 & 0.50 &  -5.70 & 3070.98 & 1.225 \\
2455784.6229 &  -8.18 & 0.48 &  -7.99 & 3075.52 & 1.371 \\
2455810.5101 &   4.61 & 0.57 &  -9.67 & 3078.15 & 1.306 \\
2455815.5504 &   6.23 & 0.55 &  -9.09 & 3072.08 & 1.562 \\
2455817.5325 &  -7.73 & 0.62 &  -7.03 & 3077.42 & 2.293 \\
2456385.7728 &  -7.98 & 0.71 &  -8.95 & 3074.62 & 1.184 \\
2456386.7932 &  -3.14 & 0.75 & -11.37 & 3069.91 & 1.241 \\
2456387.8161 &   2.25 & 0.63 &  -9.86 & 3069.89 & 1.332 \\
2456388.7910 &   7.12 & 0.57 & -10.29 & 3072.57 & 1.399 \\
2456389.7575 &   3.71 & 0.69 &  -9.42 & 3071.48 & 1.473 \\
2456390.7806 &  -9.38 & 0.85 &  -8.71 & 3074.05 & 1.581 \\
2456393.8390 &  11.11 & 0.99 &  -7.01 & 3084.10 & 1.375 \\
2456394.8345 &  -1.66 & 0.80 & -11.96 & 3080.26 & 1.331 \\
2456395.7581 &  -7.81 & 0.69 &  -8.00 & 3078.12 & 1.435 \\
2456396.8055 &  -0.72 & 0.67 & -10.36 & 3080.82 & 1.562 \\
2456397.8548 &   5.22 & 0.70 &  -8.32 & 3083.16 & 1.321 \\
2456398.7295 &   3.48 & 0.77 &  -8.76 & 3079.93 & 1.463 \\
2456399.7167 &  -7.72 & 0.68 &  -7.01 & 3090.04 & 1.446 \\
2456400.7247 &  -8.25 & 0.63 &  -7.45 & 3077.71 & 1.498 \\
2456401.7452 &   0.00 & 0.70 & -10.21 & 3080.23 & 1.647 \\
2456402.7721 &   4.74 & 0.80 &  -7.25 & 3084.09 & 1.617 \\
...\\
\hline \hline
\end{tabular}
\end{center}
\end{minipage}
\end{table*}

\begin{table*}
\caption{Example of HARPS radial velocity of GJ 674 for all 72 individual orders $v_{i}, i=1, ..., 72$.}\label{tab:HARPS_GJ674_orders}
\begin{minipage}{\textwidth}
\begin{center}
\begin{tabular}{lccccccc}
\hline \hline
$t$ [JD-2450000] & $v_{1}$ [ms$^{-1}$] & $\sigma_{v_{1}}$ [ms$^{-1}$] & $v_{2}$ [ms$^{-1}$] & $\sigma_{v_{2}}$ [ms$^{-1}$] & $v_{3}$ [ms$^{-1}$] & $\sigma_{v_{3}}$ [ms$^{-1}$] & \\
\hline
3158.7520 & -588.71 & 263.81 &   227.27 & 269.05 & 1749.74 & 276.26 & ... \\
3205.5995 &   12.31 & 101.82 &   -47.32 & 116.04 &   15.93 &  90.61 & ... \\
3237.5635 &   39.12 & 214.44 & -1319.67 & 226.17 & -159.88 & 200.46 & ... \\
3520.7932 & -134.69 & 121.59 &  -183.98 & 138.57 &  -28.22 & 105.64 & ... \\
3580.5685 &   -2.02 &  89.88 &     0.17 & 101.79 &  -39.91 &  82.02 & ... \\
3813.8477 &  -57.20 &  58.22 &  -117.54 &  69.24 &   35.22 &  50.26 & ... \\
3816.8691 & -166.60 &  65.66 &   168.84 &  77.43 &  -20.63 &  58.47 & ... \\
3861.8085 &  111.50 &  66.70 &   -93.26 &  75.73 &   25.76 &  57.69 & ... \\
3862.7863 &  -41.33 &  81.56 &   194.86 &  92.34 &   50.25 &  72.07 & ... \\
3863.8105 &   16.53 &  72.25 &    34.94 &  82.14 &   93.66 &  64.76 & ... \\
3864.7669 &  190.33 &  73.11 &    20.47 &  83.28 & -112.01 &  65.66 & ... \\
3865.8114 &  -97.21 &  63.40 &   -91.95 &  74.01 &   31.54 &  54.87 & ... \\
3866.7558 &   70.73 &  55.05 &   106.57 &  61.28 &   15.81 &  47.47 & ... \\
3867.8481 &    3.60 &  82.11 &  -120.05 &  87.68 &  -36.68 &  66.88 & ... \\
3868.8393 &  -32.43 &  68.11 &   -25.00 &  75.97 &  -88.54 &  60.51 & ... \\
3869.8038 &   22.40 &  71.35 &   135.15 &  79.57 & -183.62 &  60.78 & ... \\
3870.7388 & -225.05 & 102.91 &   -48.89 & 112.87 &   93.43 &  90.79 & ... \\
3871.8606 &   84.78 &  75.72 &   272.50 &  84.60 &    0.74 &  64.68 & ... \\
3882.7445 & -142.07 &  55.19 &    -6.12 &  63.29 &   -1.25 &  47.93 & ... \\
3886.7489 &   54.69 &  51.51 &   -34.08 &  58.77 &  -38.51 &  44.38 & ... \\
3887.7915 &   41.26 &  54.68 &   -69.51 &  62.46 &   43.17 &  47.53 & ... \\
3917.7229 &  -67.53 & 102.54 &  -307.72 & 113.21 & -164.23 &  88.00 & ... \\
3919.7271 &  266.24 & 139.73 &   138.67 & 153.85 &  118.44 & 121.60 & ... \\
3921.6271 &   55.45 &  55.68 &  -133.81 &  65.64 &  -50.26 &  51.47 & ... \\
3944.5860 & -116.93 & 111.85 &  -169.81 & 119.60 & -102.59 & 102.00 & ... \\
3947.5906 & -419.66 & 215.97 &  -574.18 & 224.53 & -506.57 & 208.66 & ... \\
3950.6474 &  -22.58 &  87.82 &   264.96 &  98.07 &  -43.89 &  77.73 & ... \\
3976.5244 &   36.70 &  65.73 &  -172.51 &  71.14 &    8.10 &  55.64 & ... \\
3980.5691 &  -66.00 & 120.14 &  -104.77 & 130.93 &  -49.01 & 104.40 & ... \\
3981.5927 &  -24.58 &  86.43 &  -118.72 &  96.98 & -119.14 &  76.33 & ... \\
3982.6162 & -150.55 &  85.05 &  -208.65 &  97.15 &   50.91 &  76.92 & ... \\
3983.5287 & -394.07 & 119.96 &   387.17 & 132.16 &  219.03 & 103.19 & ... \\
4167.8783 &   47.99 &  71.05 &    89.18 &  83.42 &   29.21 &  64.07 & ... \\
4169.8879 &  266.39 &  66.32 &    73.14 &  73.76 & -145.35 &  56.81 & ... \\
4171.8966 &   13.84 &  64.57 &   -41.04 &  77.54 &   22.22 &  57.44 & ... \\
4173.8674 & -139.35 &  71.71 &   -51.11 &  83.35 &  -22.61 &  63.72 & ... \\
4340.6256 &  -86.51 &  64.88 &   156.99 &  72.66 &   31.53 &  56.34 & ... \\
4342.5927 & -168.33 &  84.39 &    76.99 &  92.49 &  -22.83 &  77.75 & ... \\
4347.5335 &  -43.39 &  87.77 &  -237.71 & 100.36 &    6.53 &  79.37 & ... \\
4349.6053 & -288.70 & 105.74 &  -242.57 & 114.94 &  -88.61 &  92.15 & ... \\
4388.4992 &  410.95 & 131.62 &  -165.98 & 142.65 & -372.56 & 113.86 & ... \\
4589.8983 &   52.11 &  70.05 &   -91.71 &  82.61 &  193.71 &  63.27 & ... \\
4666.7202 &  159.42 & 107.67 &   117.36 & 118.26 &  -19.54 &  96.35 & ... \\
4732.4793 &  172.13 & 103.85 &   -16.30 & 116.38 & -135.62 &  89.29 & ... \\
5784.6229 &  106.28 & 140.50 &    89.25 & 151.05 &  -55.52 & 118.74 & ... \\
5810.5101 &  -43.88 & 111.11 &   -64.36 & 118.18 &  268.38 &  95.76 & ... \\
5815.5504 & -416.05 & 127.38 &  -204.83 & 140.64 &   24.47 & 107.59 & ... \\
5817.5325 & -371.97 & 129.16 &  -368.98 & 128.88 & -296.97 & 108.19 & ... \\
6385.7728 &  -26.32 & 123.76 &    26.87 & 144.05 &  109.07 & 117.45 & ... \\
6386.7932 & -116.45 & 165.94 &  -427.02 & 173.78 &  247.43 & 151.03 & ... \\
6387.8161 &  382.30 & 139.11 &  -481.63 & 149.42 &  207.95 & 115.49 & ... \\
6388.7910 &   31.58 & 142.91 &  -389.56 & 155.37 &  206.52 & 125.34 & ... \\
6389.7575 &  121.32 & 162.90 &  -384.28 & 171.63 &  -52.14 & 142.02 & ... \\
6390.7806 &  -53.28 & 184.35 &  -725.17 & 188.55 &  101.97 & 164.66 & ... \\
6393.8390 & -569.83 & 175.81 & -1284.30 & 181.88 & -152.91 & 156.36 & ... \\
6394.8345 &  -35.35 & 145.96 &  -726.20 & 161.48 &  137.19 & 126.25 & ... \\
6395.7581 & -118.25 & 141.99 & -1190.94 & 151.95 & -266.51 & 124.99 & ... \\
6396.8055 &    0.45 & 151.23 &  -983.89 & 160.24 &  -98.54 & 130.24 & ... \\
6397.8548 &  761.62 & 154.42 & -1555.73 & 160.96 & -347.23 & 132.46 & ... \\
6398.7295 &  683.26 & 154.21 &  -328.82 & 167.12 & -239.14 & 137.56 & ... \\
6399.7167 &  145.08 & 133.42 &  -315.28 & 147.14 &  101.25 & 112.60 & ... \\
6400.7247 &  669.04 & 166.52 & -1103.21 & 179.43 & -567.21 & 146.99 & ... \\
6401.7452 & -120.21 & 150.96 &  -737.78 & 162.96 & -631.43 & 129.01 & ... \\
6402.7721 &  326.55 & 179.47 & -1694.13 & 186.14 & -375.28 & 159.66 & ... \\
... \\
\hline \hline
\end{tabular}
\end{center}
\end{minipage}
\end{table*}

\begin{table}
\caption{Example of ASAS North photometry of AD Leo.}\label{tab:HARPS_GJ388_ASASN}
\begin{minipage}{0.49\textwidth}
\begin{center}
\begin{tabular}{lc}
\hline \hline
$t$ [JD] & $V$ [mag] \\
\hline
2454049.14 & 9296 \\
2454057.14 & 9277 \\
2454060.09 & 9284 \\
2454067.12 & 9285 \\
2454076.03 & 9301 \\
2454082.14 & 9245 \\
2454084.15 & 9286 \\
2454089.10 & 9283 \\
2454093.07 & 9307 \\
2454096.05 & 9293 \\
2454100.01 & 9289 \\
2454102.03 & 9269 \\
2454137.00 & 9272 \\
2454138.93 & 9284 \\
2454144.14 & 9261 \\
2454146.06 & 9271 \\
2454148.06 & 9284 \\
2454156.90 & 9283 \\
2454158.01 & 9276 \\
2454163.96 & 9292 \\
2454166.87 & 9276 \\
2454180.96 & 9281 \\
2454182.85 & 9291 \\
2454184.01 & 9300 \\
2454196.87 & 9252 \\
2454198.87 & 9290 \\
2454199.85 & 9282 \\
2454200.87 & 9287 \\
2454202.87 & 9292 \\
2454208.87 & 9288 \\
2454210.85 & 9276 \\
2454211.91 & 9290 \\
2454212.86 & 9285 \\
2454218.86 & 9295 \\
2454220.90 & 9292 \\
2454220.91 & 9324 \\
2454222.87 & 9292 \\
2454224.87 & 9291 \\
2454227.84 & 9303 \\
2454230.84 & 9271 \\
2454233.84 & 9286 \\
2454235.83 & 9262 \\
2454237.84 & 9271 \\
2454239.80 & 9282 \\
2454274.77 & 9297 \\
2454474.00 & 9282 \\
2454477.05 & 9292 \\
2454477.05 & 9299 \\
2454478.98 & 9300 \\
2454523.96 & 9273 \\
2454524.00 & 9279 \\
2454524.00 & 9274 \\
2454526.01 & 9281 \\
2454530.97 & 9274 \\
2454559.85 & 9289 \\
2454559.85 & 9271 \\
2454561.85 & 9284 \\
2454561.85 & 9282 \\
2454563.97 & 9268 \\
2454582.89 & 9260 \\
2454582.90 & 9255 \\
2454589.84 & 9263 \\
2454589.87 & 9295 \\
2454591.86 & 9248 \\
... \\
\hline \hline
\end{tabular}
\end{center}
\end{minipage}
\end{table}

\begin{table}
\caption{Example of ASAS South photometry of AD Leo.}\label{tab:HARPS_GJ388_ASASS}
\begin{minipage}{0.49\textwidth}
\begin{center}
\begin{tabular}{lc}
\hline \hline
$t$ [JD] & $V$ [mag] \\
\hline
2452622.83 & 9370 \\
2452625.83 & 9347 \\
2452628.84 & 9362 \\
2452628.84 & 9356 \\
2452635.79 & 9342 \\
2452635.79 & 9329 \\
2452637.85 & 9353 \\
2452643.80 & 9351 \\
2452645.82 & 9369 \\
2452649.78 & 9324 \\
2452651.77 & 9342 \\
2452657.77 & 9346 \\
2452661.79 & 9365 \\
2452663.78 & 9359 \\
2452665.77 & 9357 \\
2452667.77 & 9343 \\
2452669.78 & 9391 \\
2452671.76 & 9377 \\
2452673.76 & 9341 \\
2452673.76 & 9335 \\
2452675.74 & 9343 \\
2452677.73 & 9368 \\
2452681.71 & 9358 \\
2452683.71 & 9365 \\
2452690.70 & 9336 \\
2452691.81 & 9356 \\
2452694.66 & 9355 \\
2452695.71 & 9342 \\
2452696.69 & 9357 \\
2452698.68 & 9363 \\
2452700.68 & 9366 \\
2452702.67 & 9372 \\
2452704.67 & 9386 \\
2452706.66 & 9354 \\
2452710.64 & 9371 \\
2452712.62 & 9361 \\
2452717.61 & 9332 \\
2452719.68 & 9366 \\
2452723.63 & 9364 \\
2452725.63 & 9366 \\
2452727.62 & 9350 \\
2452729.62 & 9367 \\
2452733.61 & 9360 \\
2452735.60 & 9337 \\
2452737.59 & 9354 \\
2452743.57 & 9375 \\
2452745.57 & 9299 \\
2452751.54 & 9382 \\
2452754.54 & 9361 \\
2452764.53 & 9360 \\
2452776.55 & 9348 \\
2452784.49 & 9357 \\
2452786.49 & 9346 \\
2452789.48 & 9372 \\
2452811.46 & 9351 \\
2452813.47 & 9371 \\
2452975.84 & 9349 \\
2452976.85 & 9324 \\
2452978.85 & 9337 \\
2452982.83 & 9357 \\
2452994.82 & 9354 \\
2452997.83 & 9348 \\
2453005.79 & 9348 \\
2453007.82 & 9338 \\
... \\
\hline \hline
\end{tabular}
\end{center}
\end{minipage}
\end{table}


\label{lastpage}

\end{document}